\documentclass[fleqn,usenatbib]{mnras}

\usepackage{newtxtext,newtxmath}

\usepackage[T1]{fontenc}

\DeclareRobustCommand{\VAN}[3]{#2}
\let\VANthebibliography\thebibliography
\def\thebibliography{\DeclareRobustCommand{\VAN}[3]{##3}\VANthebibliography}

\usepackage{graphicx}	
\usepackage{amsmath}	
\usepackage{booktabs, makecell}

\usepackage[skip=0.333\baselineskip]{caption}
\usepackage{siunitx}
\usepackage{longtable}
\usepackage{graphicx}	
\usepackage{amsmath}	
\usepackage{algorithm}
\usepackage{algpseudocode}
\usepackage{hyperref}   
\usepackage{url} 

\title[Graph neural networks for exoplanet atmospheres]{Graph neural networks for exoplanet atmospheres}

\author[A. Vojtekova et al.]{
Antonia Vojtekova$^{1,2}$\thanks{E-mail: antonia.vojtekova.22@ucl.ac.uk},
Kai Hou Yip$^{3}$,
Ingo P. Waldmann$^{1}$
Nikolaos Nikolaou$^{1}$,
\newauthor
Olivia Venot$^{4}$,
Ahmed Faris Al-Refaie$^{1}$,
Bruno Merín$^{2}$\\
$^{1}$Department of Physics and Astronomy, University College London, Gower Street, WC1E 6BT London, United Kingdom\\
$^{2}$European Space Agency, ESAC, Camino Bajo del Castillo, 28692, Villanueva de la Cañada, Madrid, Spain\\
$^{3}$Department of Physics, King’s College London, University of London, Strand, London, WC2R 2LS, United Kingdom \\
$^{4}$Universit\'e Paris Cit\'e and Univ Paris Est Creteil, CNRS, LISA, F-75013 Paris, France \\
}

\date{Accepted XXX. Received YYY; in original form ZZZ}

\pubyear{2026}

\begin{document}
\label{firstpage}
\pagerange{\pageref{firstpage}--\pageref{lastpage}}
\maketitle

\begin{abstract}
Calculating disequilibrium chemistry in exoplanet atmospheres remains a significant computational bottleneck in atmospheric retrievals. The increasing observational precision provided by facilities such as JWST and the Ariel mission necessitates the inclusion of disequilibrium chemistry in these analyses. Previous studies have demonstrated that neural networks can emulate kinetic chemistry, although their spatial inductive bias does not align with the topology of chemical reaction networks. This study introduces a graph neural network surrogate that represents chemical species as nodes and temperature-dependent reaction rates as edges, thereby enabling information propagation along physically meaningful chemical pathways. The model is trained on atmospheres generated using the Venot+2020 chemical scheme and Guillot temperature-pressure profiles. The GNN accurately reconstructs disequilibrium abundances across the sampled parameter space and reduces the mean abundance error by a factor of approximately 3 compared to the previous U-Net model. When applied to transmission spectra, most predictions fall within the observational precision expected for JWST and Ariel, with only about 7\% of test atmospheres exceeding a 20 ppm mean spectral error. Performance variations are primarily observed in chemically transitional regimes near a carbon-to-oxygen ratio of one and at low temperatures. An evaluation of the boundary-case planet WASP-39b demonstrates effective performance under a moderate domain shift. Perturbation analysis indicates that disturbances propagate along chemical connectivity rather than spatial adjacency, confirming that the architecture captures the structure of reaction networks. These results suggest that GNN surrogates provide accurate and computationally efficient predictions of disequilibrium chemistry, thereby facilitating their integration into the atmospheric retrieval pipeline.

\end{abstract}




\section{Introduction}

In our previous work \citep{vojtekova2025}, we demonstrated the use of a fully convolutional neural network (U-Net, \citealt{ronneberger2015unet}) to accelerate the calculation of disequilibrium chemistry in exoplanetary atmospheres. In a subsequent study \citep{10.1093/rasti/rzag002}, we employed explainable artificial intelligence techniques to investigate how the network interprets input data. That analysis revealed a clear bias introduced by the use of data matrix representation and 2D convolution. While the network's overall performance was satisfactory, this bias may be problematic in scientific applications, as it potentially limits the model's performance, physicality, and interpretability.

Chemical schemes \citep{2012A&A...546A..43V, 2020A&A...634A..78V, 2024A&A...682A..52V} used to compute disequilibrium steady-state abundances naturally exhibit a graph structure, where species act as nodes and chemical interactions as edges. Consequently, adopting a graph-based data representation is a logical next step. This graph-based representation, in which key components are represented as nodes and their interactions as edges, is widely applicable across domains, from atoms connected by bonds to people in social networks.


Graph neural networks (GNNs) are a class of machine learning architectures designed to operate on graph-structured data, accommodating formats beyond typical grid-like inputs (e.g., images). 
 The foundational Message Passing Neural Network (MPNN) framework \citep{4700287, 10.5555/3305381.3305512} defines the core concept called message passing (MP). In this process, the target node is updated using messages gathered from its neighbouring nodes, which are connected to it by edges. Other types of GNNs that rely on the same core principle include Graph Convolutional Networks (GCNs) \citep{2016arXiv160902907K} and Graph Attention Networks (GATs) \citep{Velickovic2017GraphAN}. The PyTorch\footnote{\href{https://pytorch-geometric.readthedocs.io/en/latest/cheatsheet/gnn_cheatsheet.html}{PyTorch Geometric GNN Cheatsheet}} Python library provides an overview of different network types.

The versatility of GNNs has led to significant adoption across numerous scientific and engineering domains. In chemistry, GNNs are used to model quantum interactions (SCHNET, \citealt{10.5555/3294771.3294866}) or to predict molecular and crystal properties (MEGNet, \citealt{10385695}). Beyond their use in chemistry, GNNs have found widespread use in medicine \citep{ZHONG2023102640, VARSHNEY2023102535}, biology \citep{Bongini2023}, recommendation systems \citep{10.1007/s10115-025-02376-8}, particle physics \citep{particle_physics} or electric power systems \citep{10025850}.
Within astrophysics, GNNs were predominantly used in cosmology for gravitational lensing \citep{2023ApJ...953..178P}, the inference of galaxy halo masses \citep{2022ApJ...935...30V}, and cosmology simulations \citep{2022ApJ...935...30V}. In planetary science, GNNs have been used to explore orbital mechanics in the Solar System \citep{2023MLS&T...4d5002L}.

\section{Data}
In \cite{vojtekova2025}, we trained a fully convolutional neural network on synthetic data of exoplanet atmospheres with isothermal temperature–pressure (T–P) profiles. In contrast, the data utilised in this study are non-isothermal, using a T–P known as the Guillot profile \citep{2010A&A...520A..27G}. The process of data generation will be discussed in more detail in the following subsections. 

\subsection{Data Generation} 
Data are generated using the full Venot+2020 chemical scheme  \citep{2020A&A...634A..78V}, which contains 108 C/H/O/N-bearing chemical species and three condensed species: H$_2$O(c), CH$_4$(c), and NH$_3$(c). The network contains a total of 1906 reactions (948 reversible and 10 irreversible ones). We note that a new C/H/O/N chemical scheme is available \citep{2024A&A...682A..52V}, as well as a C/H/O/N/S one \citep{2026A&A...706A.260V}, but, for consistency with the previous study, we decided to use the same chemical scheme.

\subsubsection{Global parameters} \label{sec:global_param}
Following the methodology defined in \cite{vojtekova2025}, we firstly define ranges of parameters to generate atmospheric samples: 
\begin{itemize}
    \item \textbf{Planet parameters} \normalfont - mass, radius (Predicted by Forecaster \citealt{2017ApJ...834...17C} based on mass), and semi-major axis of the orbit.
    \item \textbf{Atmospheric parameters} \normalfont - carbon to oxygen ratio, metallicity, number of atmospheric layers, pressure grid.
    \item \textbf{Star parameters} \normalfont - radius, mass, type, temperature.
\end{itemize}
The pressure in all samples ranges from \(1 \times 10^{-6}\) bar to \(1 \times 10^{1}\) bar across the 128 atmospheric layers. Table~\ref{tab:atmosphere_parameters} summarises the selection and ranges of initial planetary parameters, utilising data generation techniques from the Ariel Data Challenge \citep{10.1093/rasti/rzad001}. In contrast with the previous study, we restricted planet mass to a smaller upper threshold and enforced a restriction on planet radius by enforcing surface gravity bounds of 4 -- 40 \(\mathrm{m\,s^{-2}}\).

\begin{table}
\centering
\begin{tabular}{lcc}
\hline
\multicolumn{3}{c}{\textbf{Planetary Properties}} \\ 
\hline
Parameter & Range / Distribution & Units / Notes \\
\hline
C/O Ratio & Uniform $(0.4 - 1.5)$ & - \\
Metallicity & Uniform $(0.5 - 100)$ & - \\
Planet Mass & Uniform $(0.3 - 3)$ & $M_J$ \\
\hline
\multicolumn{3}{c}{\textbf{Temperature--Pressure and Opacity Parameters}} \\ 
\hline
Parameter & Range / Distribution & Units / Notes \\
\hline
Irradiation Temp. $T_{\mathrm{irr}}$ & Uniform $(900 - 2500)$ & K \\
Internal Temp. $T_{\mathrm{int}}$ & Uniform $(50 - 150)$ & K \\
Visible Flux Fraction $\alpha$ & Uniform $(0.3 - 0.7)$ & - \\
Infrared Opacity $\kappa_{\mathrm{IR}}$ & LogUniform $(0.01 - 0.1)$ & - \\
Opacity Ratios $\gamma_{y}$ & LogUniform $(0.05 - 0.9)$ & $y = 1,2$ \\
\hline
\end{tabular}
\caption{Summary of sampled parameters used to generate the atmospheric dataset. Planetary properties are drawn from uniform distributions, while temperature–pressure profile parameters are sampled from uniform or log-uniform distributions.}
\label{tab:atmosphere_parameters}
\end{table}

\subsubsection{Temperature–pressure profile} \label{sec:TP_profile}
\begin{figure*}
\centering
\includegraphics[width=0.6\linewidth]{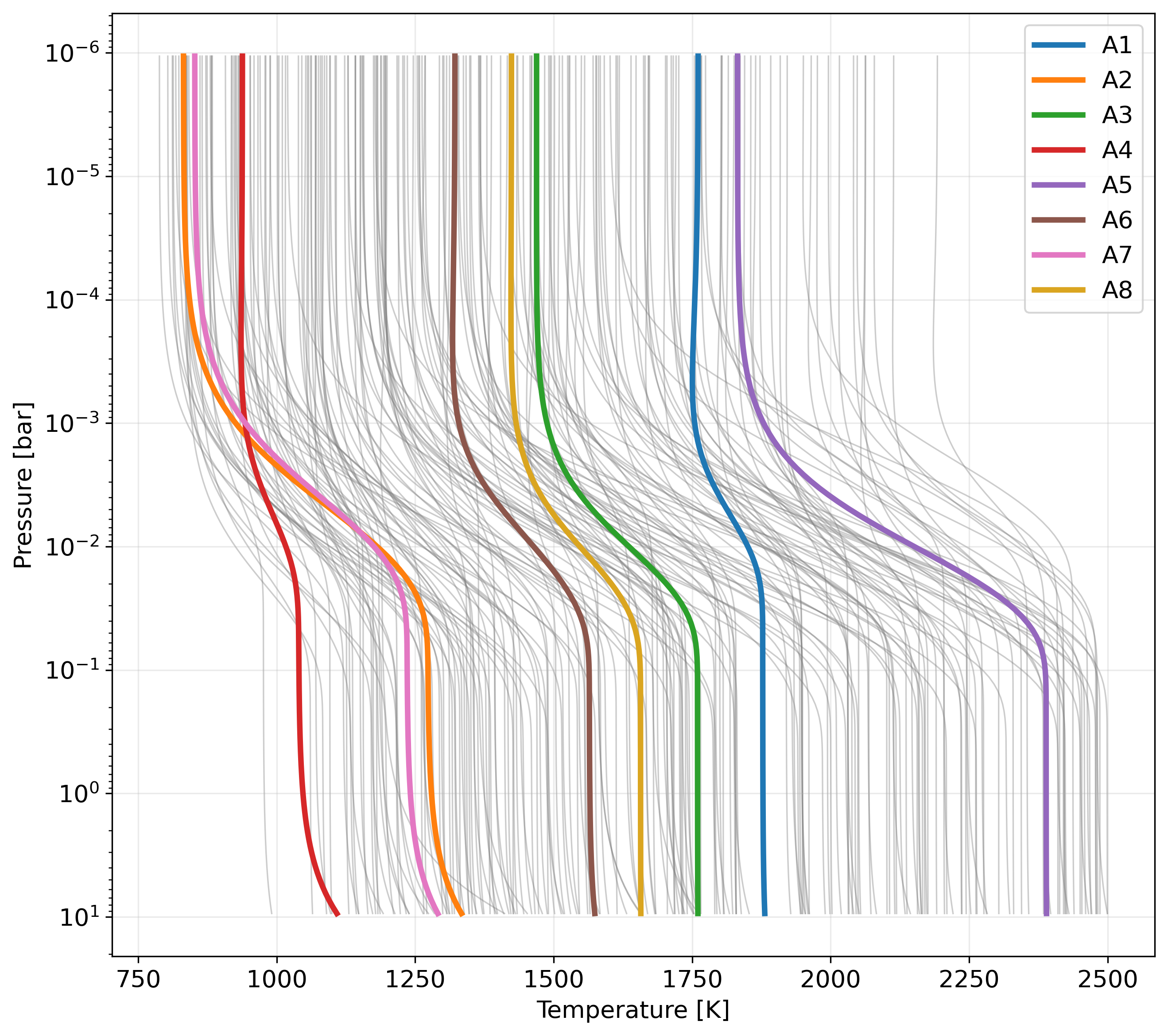} 
\caption{The Guillot temperature–pressure profiles randomly selected from the test set. The parameter range is presented in Table \ref{tab:atmosphere_parameters}. The overplotted colored profiles (A1 -  A8) correspond to the subset of samples evaluated in the Results section, with further information provided in Tables \ref{tab:gnn_samples} and \ref{tab:samples_properties}.}
\label{fig:TP_profiles}
\end{figure*}

Guillot’s T–P profile is an analytical solution to the radiative-transfer equation for irradiated, close-orbiting giant planets. The solution assumes a plane-parallel semi-grey atmosphere in radiative equilibrium, parameterised by the mean visible and infrared opacities. We implement the profile using the two-stream approximation \citep{2012ApJ...749...93L} as provided in TauREx 3.1 package \citep{2021ApJ...917...37A, 2022ApJ...932..123A}. The temperature at infrared optical depth $\tau$ is given by (Eq. 19 in \citealt{2012ApJ...749...93L}):

\begin{equation}
T^{4}(\tau) = \frac{3 T_{\mathrm{int}}^{4}}{4}\left( \frac{2}{3} + \tau \right)
+ \frac{3 T_{\mathrm{irr}}^{4}}{4} \left[ (1-\alpha)\,\xi_{\gamma_1}(\tau) + \alpha\,\xi_{\gamma_2}(\tau) \right],
\end{equation}
\begin{equation}
\xi_{\gamma_i}(\tau) = \frac{2}{3}
+ \frac{2}{3\gamma_i}\left[ 1 + \left( \frac{\gamma_i \tau}{2} - 1 \right) e^{-\gamma_i \tau} \right]
+ \frac{2\gamma_i}{3} \left( 1 - \frac{\tau^2}{2} \right) E_2(\gamma_i \tau),
\end{equation}
where $T_{\mathrm{int}}$ is the internal heat flux temperature and $T_{\mathrm{irr}}$ is the irradiation temperature derived from the incoming stellar flux. The parameters $\gamma_{1} = \kappa_{v1}/\kappa_{\mathrm{IR}}$ and $\gamma_{2} = \kappa_{v2}/\kappa_{\mathrm{IR}}$ are the ratios of visible to infrared opacities, where $\kappa_{v1}$ and $\kappa_{v2}$ are the visible opacities of the two streams and $\kappa_{\mathrm{IR}}$ is the mean infrared opacity. The parameter $\alpha \in [0,1]$ is a fraction between the two visible streams. $E_{2}(\gamma \tau)$ denotes the second–order exponential integral function.The samples are generated within the parameter ranges defined in table \ref{tab:atmosphere_parameters}.
The visible opacities are then computed as $\kappa_{vy} = \kappa_{\mathrm{IR}}\,\gamma_{y}$ with the constraint \(5\times10^{-3} \le \kappa_{vy} \le 10^{-1}\), where \(y = 1,2\). Finally, any sampled atmosphere whose temperature profile yields a maximum temperature exceeding $2500\,\mathrm{K}$ is discarded. The output of this process is a unique temperature vector with 128 values per atmospheric sample. An example of temperature–pressure profiles from the test dataset is shown in Figure \ref{fig:TP_profiles}.

\subsubsection{Equilibrium chemistry} \label{sec:EQ}
We generate equilibrium profiles following the methodology described in \cite{vojtekova2025} and utilising ACE chemistry \citep{2012A&A...548A..73A} within the TauREx package. The input parameters for the ACE chemistry and the subsequent transmission model are detailed in Table~\ref{tab:atmosphere_parameters}  and are also available in the code on GitHub\footnote{\url{https://github.com/Sponka/CHEXA-GNN}}.

\subsubsection{Disequilibrium chemistry} \label{sec:DEQ}
The last step in data generation is the calculation of disequilibrium steady-state abundances of atmospheres. As in the previous study, we use FRECKLL (Full and Reduced Exoplanet Chemical Kinetics distiLLed, \citealt{Al_Refaie_2022, 2024A&A...682A..52V}) chemical kinetic model, to include disequilibrium processes such as vertical mixing. Vertical mixing is described by the Eddy diffusion coefficient (K$_{zz}$) with a constant value for all samples equal to $10^9 \mathrm{cm}^2 \cdot \mathrm{s}^{-1}$.

\subsubsection{Reaction Rates} \label{sec:RR}
In addition to the previously used atmospheric dataset, we include reaction rate vectors for neural network training to encode a chemically meaningful connection between species. Reaction rates are usually primarily temperature-dependent, although some reactions also depend on pressure or on collisional third bodies. Reaction rates for simple temperature-dependent processes follow the modified Arrhenius law: 
\begin{equation}
    k(T) = A\,T^{n}\,\exp\left(-\frac{E_{\mathrm{a}}}{RT}\right),
\end{equation}

\noindent where \(A\) is the pre-exponential factor, \(n\) accounts for the temperature dependence of the collisional frequency, \(E_{\mathrm{a}}\) is the activation energy, \(R\) is the ideal gas constant, and \(T\) is temperature.
The code utilises the Kooij formalism, which is expressed as:
\begin{equation}
    k(T) = \alpha\,T^{\beta}\,\exp\left(-\frac{\gamma}{T}\right),
\end{equation}
where \(\gamma = E_{\mathrm{a}}/R\).

For more complex reactions that require a collisional body to proceed, such as thermal dissociation or recombination (three-body) reactions, the rate constant depends on whether it occurs under low or high pressure. Between these two limits lies a fall-off zone. There are three methods used for the representation of the rate in the fall-off zone: (1) the Lindemann mechanism \citep{TF9221700598}, (2) the Troe formulation \citep{1983JPhCh..87.1800T}, and (3) the SRI formulation \citep{STEWART198925}.

Given that the reaction rates vary with temperature (and sometimes with pressure), we extract rates individually for each atmospheric sample using the FRECKLL package. The newly released package includes built-in capabilities to easily extract and filter equations of different types. For each atmosphere, we initialised the chemical network, loaded the corresponding temperature--pressure profile and VMR abundance array, and then evaluated the reaction set using \texttt{network.compute\_reactions(vmr, temperature, pressure)}.

\subsection{Data preprocessing}
This section describes how the generated atmospheric data are transformed into inputs for the graph neural network. This process can be divided into several steps: filtering data (removing outliers and selecting target species), formatting data (correctly formatting the generated data), and normalising data (per input type).

\subsubsection{Data Filtering} \label{sec:filtering}
Three types of filtering were implemented on the generated dataset: (1) molecule selection, (2) outlier removal and interpolation and (3) temperature filtering. 

In contrast to our previous study, which included the 64 most abundant species, the present work restricts the number to 40 species. These species were selected by cross-matching the previously identified set with the reduced chemical scheme Venot+2020.

This dataset also presented challenges due to the presence of outliers. These outliers arose from the solver’s precision settings. This caused the log(VMR) to drop abruptly to -50 for particular species, causing jitter. Specifically, the affected values dropped suddenly from log(VMR) values (e.g. around -25) to  -50 and then oscillated around this limit for several subsequent points (e.g. 10-20 values at low pressures). Although decreasing the absolute tolerance could mitigate this issue, it would substantially increase computational cost. Consequently, we removed samples containing more than 10 affected points per molecule and a total of 20 affected points. For samples with fewer than 10 affected points, we replaced the values by interpolating the third last value preceding the onset of the jitter. The jitter occurred predominantly in C$_2$H$_5$, C$_2$H$_6$, C, and O$_2$, meaning that undetected cases contribute to increased errors in the trained network for these species.

Lastly, we restricted the temperature values for the T--P profiles to the range [700 K,2500 K] to remove extremely low/high temperature samples.

After filtering, the total dataset contains 22973 samples. This is split into training, validation, and test datasets at an 80/10/10 ratio. For comparison with the U-Net architecture, we use the same train/validation/test split.

\subsection{Data Formatting} \label{sec:formatting}
We distinguish several subtypes of input data, which are formatted to the three main data formats of the graph neural network: nodes, edges and global information (Table \ref{tab:parameters_summary}). The network output is a set of nodes with abundances in disequilibrium steady-state abundances.

\textit{Nodes} represent normalised VMR. Every node feature is denoted as $\mathbf{h}_i^{(l)}$, $l=0$ corresponding to the initial equilibrium abundances. Each node has a structure of 128 atmospheric (T–P) layers. The network has 40 nodes, equal to the number of species. The number of nodes and the atmospheric layers are the same for every atmospheric sample. 

\begin{figure*}
\centering
\includegraphics[width=0.49\linewidth]{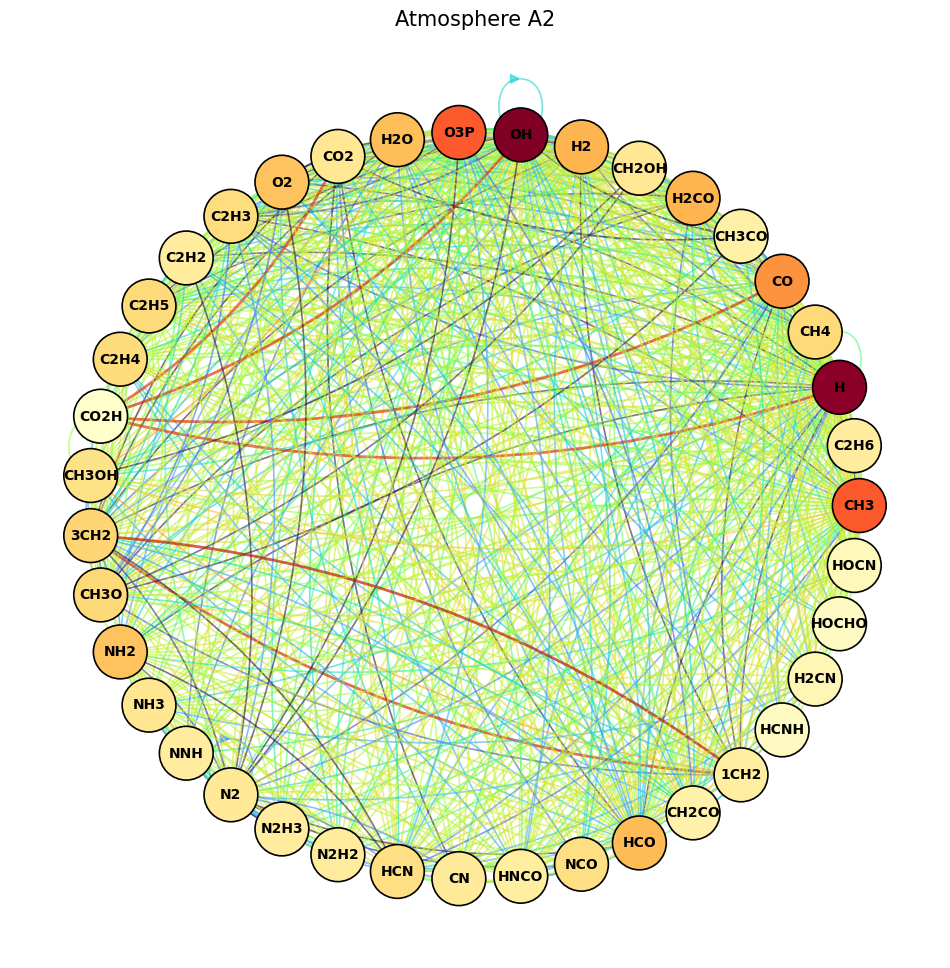} 
\includegraphics[width=0.49\linewidth]{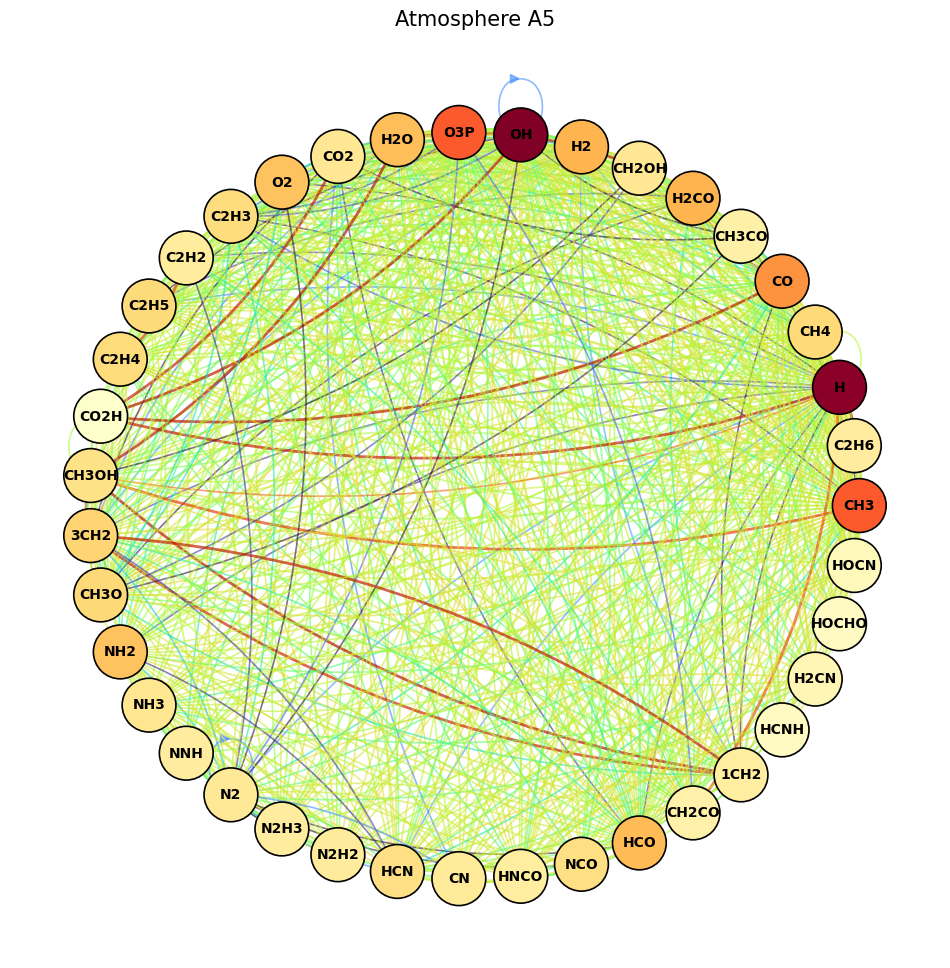} 
\includegraphics[width=0.49\linewidth]{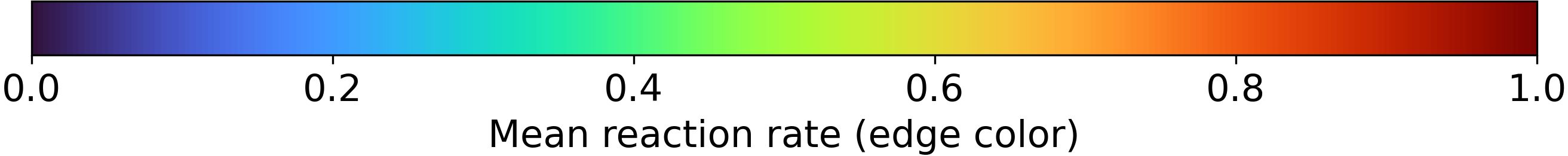} 
\includegraphics[width=0.49\linewidth]{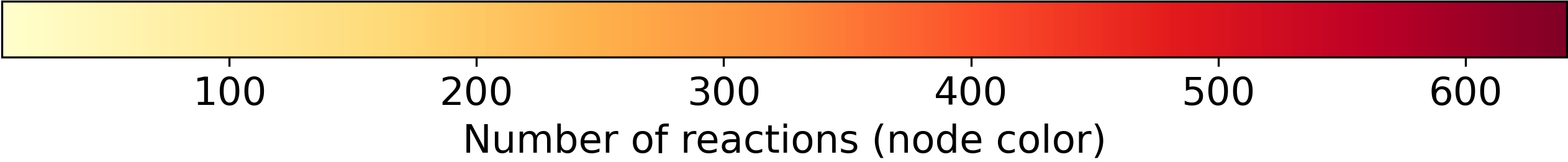} 
\caption{Visualisation of a chemical reaction network comparing two atmospheres with irradiation temperatures 945$K$ and 2015$K$ for atmospheres A2 and A5, respectively. The temperature–pressure profiles of the selected samples are visualised in Figure \ref{fig:TP_profiles}. The parameters for both atmospheres are detailed in Table \ref{tab:samples_properties}. While all species participate in the same number of reactions in both atmospheres, reaction rates (edges) differ in magnitude, as indicated by the colour scale. This highlights the effect of atmospheric temperature on reaction rates.}
\label{fig:sample_graph}
\end{figure*}

\textit{Edges} represent connections between nodes — in this study, reaction rates between species. The edges in this study are directed, representing the flow of information from reactant species to product species. Each edge is denoted as $\mathbf{k}_{ij}$ and is a vector of 128 normalised reaction rates. Every sample has the same number of edges; however, the reaction rates within each edge vary due to temperature-dependent reaction rates (and other factors). The total number of edges in the network is 3286. Figure~\ref{fig:sample_graph} shows a complete chemical reaction network for atmospheres with different initial configurations, where nodes are represented as circles with the name of the species and edges are represented as arrows between the species. Helium is the only species which has an artificially created edge, as it does not react with any other species in the network. We have added a self-loop with a value of 0.5.

\textit{Global parameters} represent the parameters of every atmospheric sample, denoted as $\mathbf{u}$. This vector contains 128 values for the temperature profile and 13 planetary parameters, for a total of 141 values. The planetary parameters include the C/O ratio, metallicity, planet mass, planet radius, the derived surface gravity, and the parameters defining the temperature–pressure profile: irradiation temperature, internal temperature, infrared opacity, two visible opacities, two opacity ratios, and the visible flux ratio. The values of $\mathbf{u}$ change for each atmospheric sample, but the dimensionality remains fixed. 

\subsubsection{Data Normalisation} \label{sec:normalisation}

The abundances, reaction rates, and global parameters span many orders of magnitude. Logarithmic transformations are applied only to abundances and reaction rates; all global parameters remain in linear space.

\textbf{Molecular abundances (node features).} Both the equilibrium inputs and the disequilibrium steady-state abundances are initially in units of volume mixing ratios (VMRs). The first step is to log-transform the data, and clip very low abundances to a fixed lower bound of $-30$. Afterwards, the values are linearly rescaled to range $[-1,1]$.

\textbf{Reaction rates (edge features).} Each reaction rate coefficient $k$ is first log-transformed, and to avoid extreme outlier values, the data are clipped
\[
r = \log_{10}(k),
\qquad
\mathrm{clip}(r;\, r_{\min}, r_{\max}) =
\begin{cases}
r_{\min}, & r < r_{\min},\\[3pt]
r_{\max}, & r > r_{\max},\\[3pt]
r, & \text{otherwise},
\end{cases}\]
where $r_{\min} = -40$ and $r_{\max}=10.1$. The resulting values are normalised to the range $[0,1]$. We additionally restrict normalised reaction rates to the interval $[10^{-5},\,0.9999]$ to avoid exact zeros or ones in the edge encoder.

\textbf{Global parameters.} The global input vector consists of planetary and atmospheric properties as shown in Table \ref{tab:atmosphere_parameters}. Each parameter is min--max scaled independently to the range $[-1,1]$ using bounds relevant to each parameter.

\section{Methods}
This study explores the use of a message-passing graph neural network. The goal of the selected architecture is to create a surrogate model to calculate disequilibrium steady-state abundances from input equilibrium abundances. 

\begin{table*}
\centering
\begin{tabular}{l l c c p{6cm}}
\hline
\textbf{Name} & \textbf{Architecture Part} & \textbf{Input Size}  & \textbf{Symbol} & \textbf{Note} \\
\hline
Nodes 
& Node Encoder / MPB
& 128 
& $\mathbf{h}_i^{(l)}$ 
& Initial node data ($\mathbf{h}_i^{(0)}$) correspond to equilibrium abundances (Section~\ref{sec:EQ}). \\[3pt]

Edges 
& Edge Encoder / MPB
& 128   
& $\mathbf{k}_{ij}$ 
& Directed reaction-rate vectors from reactant to product (Section~\ref{sec:RR}). \\[3pt]

Message 
& Message Passing Block (MPB)
& 128  
& $\mathbf{m}_i$ 
& Aggregated neighbour information weighted by edge features. \\[3pt]

Global parameters 
& Global Encoder / Node Encoder 
& 141 
& $\mathbf{u}$ 
& Temperature profile and planetary parameters (Section~\ref{sec:global_param}). \\[3pt]

\hline
\end{tabular}
\caption{Summary of the main data components used in the graph architecture. Formatting of the data is described in Section~\ref{sec:formatting}, data filtering in Section~\ref{sec:filtering} and data normalisation in Section~\ref{sec:normalisation}.}
\label{tab:parameters_summary}
\end{table*}

\begin{table*}
\centering
\begin{tabular}{l c c c c c c}
\hline
\textbf{Name} & \textbf{Input Size} & \textbf{Hidden dimension}  & \textbf{Output Size} & \textbf{Number of Linear Layers}  & \textbf{Learnable parameters}  & \textbf{Figure} \\
\hline
Node Encoder 
& 384
& H$_{\text{NE-MLP}} = 512$ 
& 128 
& 4
& \num{788096}
& Fig. \ref{fig:GNN-NE-MLP}
 \\

Edge Encoder 
& 128
& H$_{\text{EE-MLP}} = 256$   
& 128 
& 3
& \num{131968}
& Fig. \ref{fig:GNN-EE}
 \\

Global Encoder 
& 141 
& H$_{\text{GE-MLP}} = 256$  
& 128 
& 3
& \num{135040}
& Fig. \ref{fig:GNN-GE-MLP}
 \\

Decoder 
& 128 
& H$_{\text{D-MLP}} = 1024$  
& 128 
& 2
& \num{263296}
& Fig. \ref{fig:GNN-D-MLP}
 \\

\hline
\end{tabular}
\caption{Summary of network module architectures and hyperparameters. For each MLP module, the table reports the input dimensionality, hidden-layer size, output dimensionality, number of linear layers, total number of learnable parameters, and the corresponding schematic figure. All modules employ a Leaky ReLU activation function ($a = 0.2$). With the exception of the decoder, the full GNN architecture contains four instances of each module.}
\label{tab:Network_summary}
\end{table*}

\subsection{Model Architecture}

\begin{figure*}
\centering
\includegraphics[width=\linewidth]{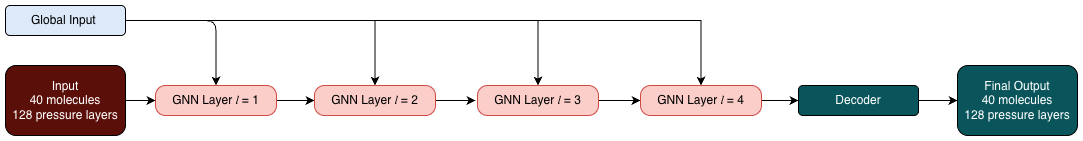} 
\caption{Overview of the GNN. The network receives as input the normalised abundances at thermochemical equilibrium and the global parameters of the target atmosphere, and outputs the disequilibrium steady-state abundance profile. Data passes through four graph neural network modules followed by a final decoder. Hyperparameters of the network modules are provided in Table \ref{tab:Network_summary}.}
\label{fig:GNN-Schematic}
\end{figure*}

The graph network (Figure~\ref{fig:GNN-Schematic}) is constructed from several smaller blocks (Figures \ref{fig:GNN-layer} -- \ref{fig:GNN-NE-MLP}), namely (1) graph layers (Figure~\ref{fig:GNN-layer}) and (2) a decoder (Figure~\ref{fig:GNN-D-MLP}). The graph layers are further constructed from four modules: (A) an edge-encoding MLP (Figure~\ref{fig:GNN-EE}), (B) a message-passing block (Figure~\ref{fig:GNN-MA}), (C) a global encoder MLP (Figure~\ref{fig:GNN-GE-MLP}) and (D) a node-encoder MLP (Figure~\ref{fig:GNN-NE-MLP}). Every graph layer contains a unique set of (A) – (D) MLPs. The complete architecture contains one decoder, and $L$ graph layers. The summary of network components and dimensions for all MLPs is in Table \ref{tab:Network_summary}. A comprehensive explanation of the data can be found in Section 2 and summarised in Table \ref{tab:parameters_summary}.

The essential idea behind the architecture is updating node information (species abundance) based on information (message) from neighbouring nodes (species) which react together. 

This section describes distinct modules, firstly the singular modules (one per complete graph network) and afterwards the graph layer architecture and the modules it contains.

Several architectural components are repeated across every module. First is the multilayer perceptron (MLP). An MLP is a feed-forward neural network that takes an input $\mathbf{x} = \mathbf{h}_i^{(0)}$ and produces an output $\mathbf{y}$. The input is transformed through a sequence of $L$ hidden linear layers and non-linear activations, enabling the network to learn complex, non-linear relationships. This can be expressed as:

\begin{equation}
    \mathbf{h}_i^{(l)} = f\!\left( \mathbf{W}^{(l)} \, \mathbf{h}_i^{(l-1)} + \mathbf{b}^{(l)} \right), \qquad l = 1, \dots, L-1,
\end{equation}

\begin{equation}
    \mathbf{y}_i = \mathbf{W}^{(L)} \, \mathbf{h}_i^{(L-1)} + \mathbf{b}^{(L)} ,
\end{equation}

\noindent where $\mathbf{x} = \mathbf{h}_i^{(0)}$ is the input feature of node $i$, $\mathbf{W}^{(l)}$ and $\mathbf{b}^{(l)}$ are the learnable weights and biases of layer $l$, $f(\cdot)$ is the non-linear activation function, and $\mathbf{y}_i$ is the final output for node $i$. 
The second shared component is the Leaky ReLU activation function, defined as:
\begin{equation}
    f(z) = \begin{cases} z, & z \ge 0, \\  a z, & z < 0, \end{cases}
\end{equation}
where $a$ is the negative slope parameter, set to $0.2$ in this study.
The next shared component is layer normalisation \citep{2016arXiv160706450L}, which is defined as:
\begin{equation}
    \mathbf{y} = \frac{ \mathbf{x} - \boldsymbol{\mu} } { \sqrt{ \boldsymbol{\sigma}^2 + \varepsilon } }  \odot \boldsymbol{\gamma} + \boldsymbol{\beta},
\end{equation}
where $\mathbf{x}$ is the input vector,  $\boldsymbol{\mu}$ and $\boldsymbol{\sigma}^2$ are the mean and variance computed over the features of $\mathbf{x}$, $\varepsilon$ is a small numerical constant for stability, $\boldsymbol{\gamma}$ and $\boldsymbol{\beta}$ are learnable scaling and shifting parameters, and $\mathbf{y}$ is the normalised output. Symbol $\odot$ denotes elementwise multiplication. This layer enables learnable normalisation parameters, helping stabilise network training and avoid exploding gradients.

\subsubsection{Decoder MLP}
The final transformation of the network is performed by the shared decoder (Figure~\ref{fig:GNN-D-MLP}), which includes two linear layers and one activation function to ensure consistency across species. The input to the decoder is the final node representation $\mathbf{h}_i^{(L)}$ of size $128$, the hidden layer contains $1024$ units, and the output is the decoded vector $\mathbf{y}_i$ of size $128$ for each species.

\subsubsection{Graph Neural Network Layer}
\begin{figure*}
\centering
\includegraphics[width=0.6\linewidth]{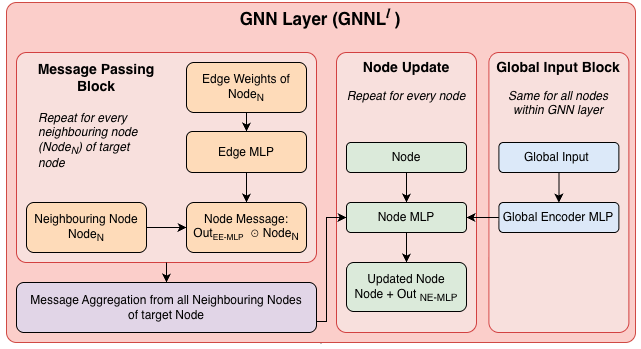} 
\caption{Schematic visualisation of a single graph neural network layer composed of three modular blocks: global encoding, message passing, and node update. Global information is first processed by the global encoder MLP, which is evaluated once per GNN layer and shared across all nodes. In parallel, message passing is performed for each target node by processing its neighbouring nodes (Node$_N$) through the edge encoder MLP. For each neighbour, the resulting edge message is weighted by the normalised abundance of the corresponding neighbouring node. Messages from all neighbours are then aggregated to form a single message for the target node. This procedure is repeated independently for every node in the graph. Finally, the aggregated message, the global encoding, and the target node features are combined and passed through the node encoder MLP. The resulting output is added residually to the target node features to produce the updated node representation. The output of the final GNN layer is subsequently passed to the decoder to generate the final prediction.}
\label{fig:GNN-layer}
\end{figure*}

\begin{figure}
\centering
\includegraphics[width=0.7\linewidth]{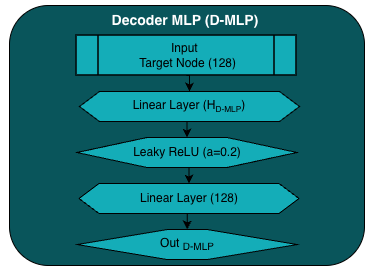} 
\caption{Schematic of the decoder MLP. The decoder receives updated target node representations as input and applies a series of linear transformations, followed by non-linear activations, to generate the final output. Each node is processed independently.}
\label{fig:GNN-D-MLP}
\end{figure}

The most complex component of the architecture is the GNN Layer ($\mathrm{GNNL}^{(l)}$, Figure~\ref{fig:GNN-layer}), which consists of three separate MLPs and a message passing module. There are $L$ graph layers in total, each with its own set of learnable parameters.

The data flow through each graph layer as follows: First, a \emph{message} $\mathbf{m}_i^{(l)}$ is computed by aggregating information from all neighbouring nodes $\mathbf{h}_j^{(l)}$, weighted by the corresponding edge attributes. These edge attributes are processed by the \emph{edge encoder} MLP, which is shared across all edges within the graph layer.

To update a target node, the aggregated message $\mathbf{m}_i^{(l)}$, the current node features $\mathbf{h}_i^{(l)}$, and the global encoder output $\mathbf{u}'$ are concatenated. This concatenated vector is passed through the \emph{node encoder} MLP, which is shared across all species within the same graph layer. Finally, the node representation is updated using a residual connection between the original node feature $\mathbf{h}_i^{(l)}$ and the output of the node encoder MLP, yielding the updated representation $\mathbf{h}_i^{(l+1)}$. This residual connection mitigates oversmoothing, a common issue in deeper graph architectures.

\begin{figure}
\centering
\includegraphics[width=0.7\linewidth]{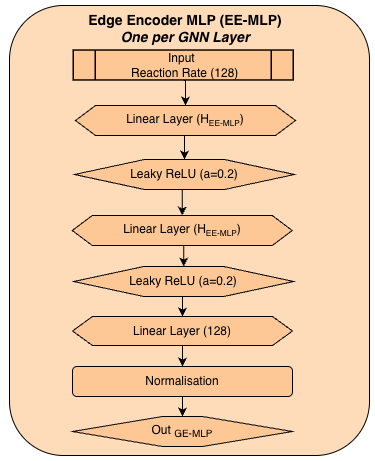} 
\caption{Schematic of the edge encoder MLP. For each GNN layer, the edge encoder MLP processes edge features corresponding to reaction rates. Each edge connecting a target node to one of its neighbouring nodes is handled independently by the MLP.}
\label{fig:GNN-EE}
\end{figure}

\textbf{(A) Edge Encoder MLP ($\mathrm{MLP}_{\mathrm{EE}}$)}. 
The purpose of this network is to process the reaction rate associated with each pair of neighbouring nodes. The module consists of two linear layers with two activation functions, followed by a final linear layer and a layer normalisation step (Figure~\ref{fig:GNN-EE}). 

The input is a normalised reaction-rate vector
\begin{equation}
    \mathbf{k}_{ij} = k_{ij}(\mathbf{T}) \quad 
    \text{(reaction-rate vector with 128 temperature points)}, 
\end{equation}
which is passed through the edge-encoding network,
\begin{equation}
    \mathbf{e}_{ij} = \mathrm{MLP}_{\mathrm{EE}}(\mathbf{k}_{ij}),
\end{equation}
producing the edge feature $\mathbf{e}_{ij}$. This feature is subsequently used in the Message Passing Block.

\begin{figure}
\centering
\includegraphics[width=0.7\linewidth]{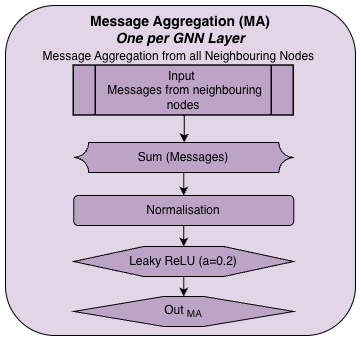} 
\caption{Message Aggregation module schematic. For each GNN layer, the Message Aggregation module collects all messages from neighbouring nodes of a given target node and aggregates them. The aggregated message is subsequently normalised to prevent parameter explosion and passed through a non-linear activation. All neighbour messages associated with a target node are processed jointly.}
\label{fig:GNN-MA}
\end{figure}

\textbf{(B) Message Passing Block ($MPB$)}.
The goal of this block is to aggregate messages from all neighbouring nodes of a target node (Figures ~\ref{fig:GNN-layer} and ~\ref{fig:GNN-MA}). Neighbours are defined by the presence of a chemical reaction connecting the corresponding species.
The aggregated message for node $i$ at layer $l$ is given by
\begin{equation}
    \mathbf{m}_i^{(l)} = \sum_{j \in \mathcal{N}(i)} \mathbf{m}_{ij}^{(l)},
\end{equation}
where $\mathbf{m}_{ij}^{(l)}$ is the message contributed by neighbour $j$, computed as
\begin{equation}
    \mathbf{m}_{ij}^{(l)} = \mathrm{MLP}_{\mathrm{EE}}(\mathbf{k}_{ij}) \odot \mathbf{h}_j^{(l)},
\end{equation}
and $\odot$ denotes element-wise multiplication. The resulting message $\mathbf{m}_i^{(l)}$ (vector of 128 values) is then used as input to the node encoder MLP.

\begin{figure}
\centering
\includegraphics[width=0.7\linewidth]{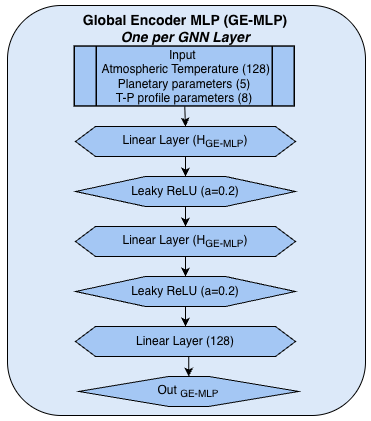} 
\caption{Global encoder MLP schematic. For each GNN layer, the global encoder MLP processes global atmospheric information. The resulting embedding is shared across all nodes within the GNN layer and provided as additional contextual input to the node encoder MLP.}
\label{fig:GNN-GE-MLP}
\end{figure}

\textbf{(C) Global Encoder MLP ($\mathrm{MLP}_{\mathrm{GE}}$)}. 
The global encoder MLP consists of three linear layers and two activation functions, as shown in Figure~\ref{fig:GNN-GE-MLP}, with the final layer being linear. The input to this network is a vector $\mathbf{u} \in \mathbb{R}^{141}$, which is transformed into $256$ hidden units, and the output is a vector $\mathbf{u}' \in \mathbb{R}^{128}$. The output $\mathbf{u}'$ of the global encoder MLP is subsequently fed into the node encoder MLP in each graph layer. Every graph network contains one global encoder MLP; in total, there are four global encoder MLPs within the full network.

\begin{figure}
\centering
\includegraphics[width=0.7\linewidth]{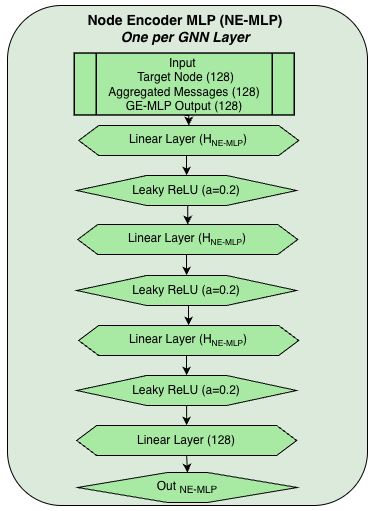} 
\caption{Node encoder MLP schematic. For each GNN layer, the node encoder MLP updates node representations by jointly processing the target node features, the aggregated message from neighbouring nodes, and the output of the global encoder. The module is applied independently to each node, incorporating both local neighbourhood information and global context.}
\label{fig:GNN-NE-MLP}
\end{figure}

\textbf{(D) Node Encoder MLP ($\mathrm{MLP}_{\mathrm{NE}}$)}.
The most crucial component of the graph layer is the node encoder MLP (Figure~\ref{fig:GNN-NE-MLP}), which takes as input the target node feature $\mathbf{h}_i^{(l)}$, the aggregated message from neighbouring nodes $\mathbf{m}_i^{(l)}$, and the processed global parameters $\mathbf{u}'$. The purpose of this module is to update the abundance of the target node.
The network consists of three pairs of linear and activation layers, followed by a final linear layer. The input vector ($\mathbf{z}_i^{(l)}$) is formed by concatenating the three components, with a total dimensionality of $384$. The hidden layers contain $512$ units, and the output layer has size $128$. A residual connection is applied to mitigate oversmoothing. 
We denote the concatenated input as
\begin{equation}
    \mathbf{z}_i^{(l)} = \text{concat} \left(\, \mathbf{h}_i^{(l)}, \mathbf{m}_i^{(l)}, \mathbf{u}' \,\right),
\end{equation}
and the node update rule becomes
\begin{equation}
    \mathbf{h}_i^{(l+1)} =  \mathbf{h}_i^{(l)} + \mathrm{MLP}_{\mathrm{NE}}\!\left( \mathbf{z}_i^{(l)} \right).
\end{equation}

\subsection{Training Strategy}

The network is trained using the Mean Absolute Error (MAE) loss function 
and the Adam optimiser. We employed a learning rate scheduler that reduced the learning rate when the validation loss plateaued, along with early stopping to prevent overfitting. Training used a batch size of 8 for up to 150 epochs. Training converged after 138 epochs and required around 10 hours on an NVIDIA L4 GPU.


\subsection{Hyperparameter Search}

We employed structured hyperparameter search to tune critical architectural parameters, including the number of linear layers and width (number of units) within the various MLPs, the number of graph layers, the type of message aggregation, the form of global parameter involvement, etc. Table \ref{tab:hyperparameter_search} contains the explored parameters in this study. Certain non-search architectural choices, such as the use of layer normalisation, were selected based on analysis of the network's internal behaviour, including the distribution of weights. For a thorough evaluation, we selected five top-performing candidates based on their validation loss. The final model was selected from this subset based on its overall performance across our evaluation metrics. Due to the large number of models trained during this process, we cannot present all experimental results. Instead, we compare our resulting final model against a previously published U-Net model \citep{vojtekova2025}. The U-net model is retrained with the same dataset as the graph neural network for direct comparison.

\subsection{Evaluation}

We use the same evaluation methodology as in our previous study \cite{vojtekova2025}, combining numerical and visual techniques. The first numerical metric is the Mean Absolute Error (MAE), defined as
\begin{equation}
\mathrm{MAE} =\frac{1}{n P S}\sum_{i=1}^{n}\sum_{p=1}^{P}\sum_{s=1}^{S} \left| y_{ips} - \hat{y}_{ips} \right|,
\end{equation}
where $n$ is the number of atmospheres, $P$ the number of pressure (atmospheric) layers, $S$ the number of species, $y_{ips}$ the ground truth value, and $\hat{y}_{ips}$ the corresponding network prediction.

Next is the Mean Absolute Percentage Error (MAPE),
\begin{equation}
\mathrm{MAPE} =\frac{100}{n P S}\sum_{i=1}^{n} \sum_{p=1}^{P} \sum_{s=1}^{S} \left| \frac{y_{ips} - \hat{y}_{ips}}{y_{ips}}\right|,
\end{equation}
which expresses the relative prediction error as a percentage of the ground truth.
A detailed discussion of these metrics is provided in \cite{vojtekova2025}, Section~2.4.

\subsection{Full forward modelling}
The surrogate model for predicting disequilibrium steady-states is evaluated by transforming abundance data into transmission spectra using TauREx. For each sample, predicted chemical abundances are combined with planetary and stellar parameters, along with a Guillot-type temperature–pressure profile, to compute synthetic transmission spectra. The forward model incorporates molecular absorption, collisionally induced absorption (CIA) from H$_2$–H$_2$ and H$_2$–He pairs, and Rayleigh scattering; clouds are excluded from the simulations.

The discrepancy is quantified using the mean absolute error (MAE) in transit depth, expressed in parts per million (ppm), between spectra generated from network-predicted abundances and those produced by the FRECKLL code (ground-truth samples). This approach provides valuable insight because transmission spectra are derived from specific atmospheric pressure layers rather than the entire atmosphere. As a result, errors in the highest and lowest pressure layers exert less influence on the spectrum than errors in mid-pressure layers. This method, therefore, offers a more realistic assessment of the network's performance in atmospheric retrieval tasks.

Although the MAE is computed on the native wavelength grid produced by TauREx, spectra shown for visualisation are binned to the spectral resolution of JWST/NIRSpec G395H. This allows a comparison with observational precision. For reference, we adopt an instrumental noise floor of 20 ppm, representative of the best-case performance expected for JWST and comparable to the target precision of Ariel.


\subsubsection{WASP-39b}
To evaluate the network under realistic atmospheric conditions and maintain continuity with previous work \citep{vojtekova2025}, WASP-39b was selected as the test case. A Guillot-type temperature–pressure profile was reconstructed to approximate the retrieved profile of \cite{Khorshid2024}, while remaining within the bounds of the training parameter space. Since the retrieved profile includes parameters outside the sampled range, the adopted parameters (Table \ref{tab:parameters_wasp39b}) provide the closest in-distribution approximation. The resulting atmosphere is positioned near the boundary of the training distribution in both temperature–pressure structure and planetary properties, thereby serving as an edge-of-distribution test for the GNN emulator.

\begin{table*}
    \centering
\begin{tabular}{cccccccccccc}
\toprule
 $R_{\mathrm{p}}$ & $M_{\mathrm{p}}$ & Met. & $\mathrm{C/O}$ & $T_{\mathrm{irr}}$ & $T_{\mathrm{int}}$ & $\alpha$ & $\kappa_{\mathrm{IR}}$ & $\kappa_{\mathrm{v1}}$ & $\kappa_{\mathrm{v2}}$ & $\gamma_1$ & $\gamma_2$ \\
$[R_{\mathrm{J}}]$ & $[M_{\mathrm{J}}]$ & & Ratio & [K] & [K] & & & & & & \\
\midrule
 $ 1.27 $ & $ 0.30 $ & $ 1.32 $ & $ 0.48 $ & $ 900 $ & $ 50 $ & $ 0.7 $ & $ 1.0 \cdot 10^{-1} $ & $ 5.0 \cdot 10^{-3} $ & $ 5.0 \cdot 10^{-3} $ & $ 5.0 \cdot 10^{-2} $ & $ 5.0 \cdot 10^{-2} $ \\
\bottomrule
\end{tabular}
\caption{Atmospheric and planetary parameters adopted for WASP-39b. Parameters follow the same definition as Table~\ref{tab:samples_properties} and correspond to the Guillot temperature--pressure profile used to approximate the retrieved atmosphere.}
\label{tab:parameters_wasp39b}
\end{table*}

\begin{figure}
\centering
\includegraphics[width=\linewidth]{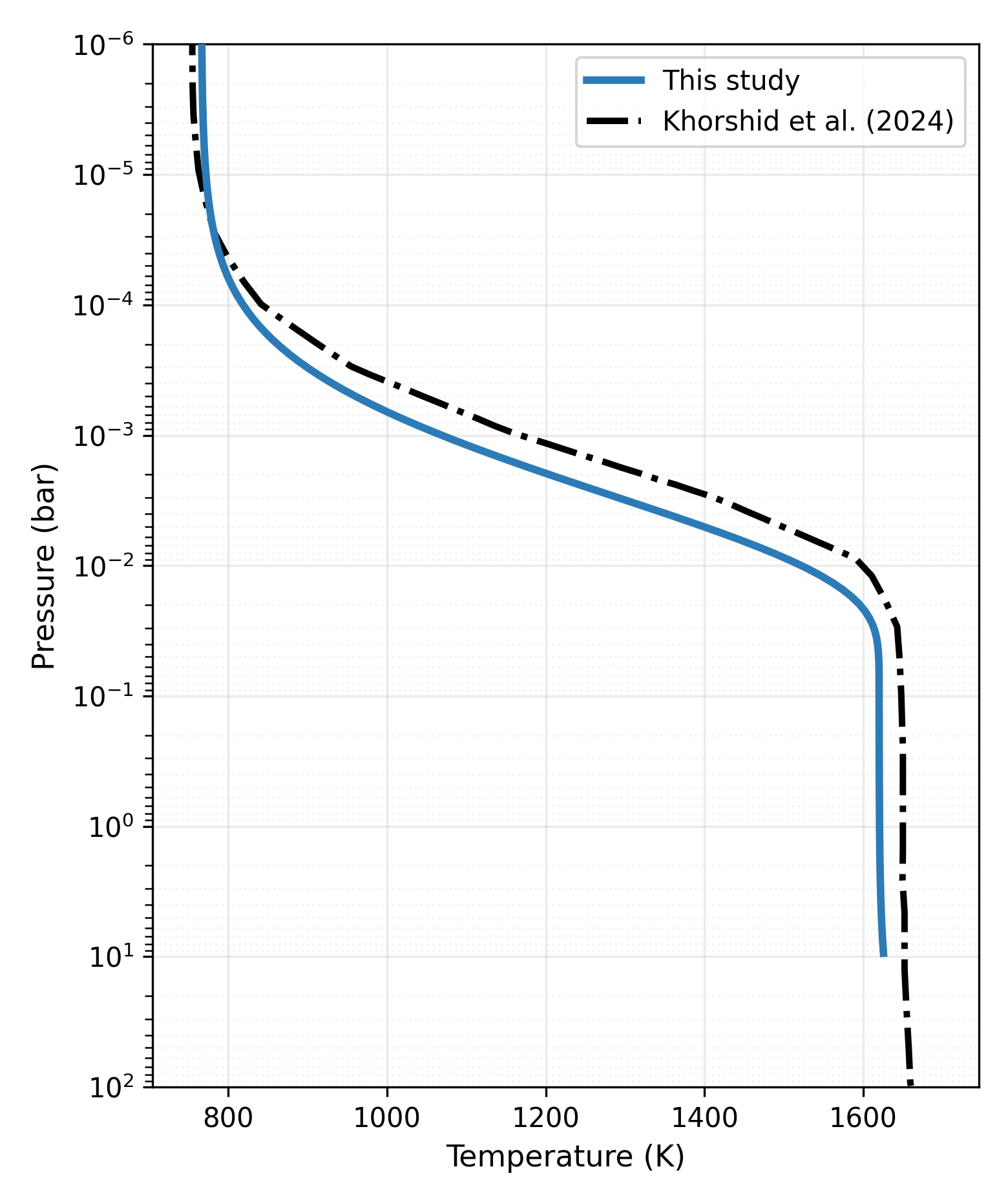} 
\caption{Temperature–pressure profile adopted for WASP-39b. The black dash–dotted curve shows the retrieved profile from \protect\cite{Khorshid2024}, while the blue curve denotes the Guillot-type approximation used in this study. The approximation was constructed to match the retrieved structure as closely as possible while remaining within the training parameter range (Table \ref{tab:atmosphere_parameters}). The corresponding Guillot parameters are listed in Table \ref{tab:parameters_wasp39b}.}
\label{fig:tp_wasp39b}
\end{figure}

\subsection{Perturbation test}

As described by \cite{10.1093/rasti/rzag002}, perturbation is an explainable artificial intelligence technique used to investigate the workings of neural networks. Previous work demonstrated that convolutional neural networks (CNNs) do not reliably learn chemically meaningful relationships between species solely from input data. In this study, a single-molecule perturbation is performed by introducing noise to a selected molecule and observing the resulting changes in output. This method is a post-hoc, model-agnostic local approach \citep{molnar2025}. 

In graph neural networks, species are represented as nodes and chemical reactions as edges. We hypothesise that perturbing an input molecule will most strongly influence those species directly connected to it. However, because our graph becomes fully connected after two steps, we expect the perturbation’s effect to propagate beyond immediate neighbours and partially influence all species in the network.

As described in \cite{10.1093/rasti/rzag002}, the perturbation is done as
\begin{equation}
    \text{Input}_{i,j}^{\text{perturbed}} =
    \begin{cases}
        \text{Input}_{i,j} + \epsilon_{i,j}, & \text{if } i = \text{target\_molecule} \\
        \text{Input}_{i,j}, & \text{otherwise}
    \end{cases}
\end{equation}
Here, \(\text{Input}_{i,j}^{\text{perturbed}}\) represents the input matrix after perturbation, where \(i\) indexes species and \(j\) indexes atmospheric layers. \(\epsilon_{i,j}\) is the perturbation term applied only to the selected molecule across all layers. In the results presented here, we apply perturbations by adding Gaussian noise with a mean of 0 and a standard deviation of 0.1. All other species remain unchanged.

In this study, we demonstrate improvements over the spatial bias inherent in the U-Net architecture; however, an in-depth investigation of perturbation effects is beyond the scope of this work.

\section{Results}
This section evaluates the performance of the proposed graph neural network and compares it to the modified U-Net architecture previously trained and analysed in \cite{vojtekova2025}. The motivation for surrogate models stems from the computational cost of the kinetic solver, which requires approximately 2 to 5 minutes per atmospheric sample, whereas the GNN reduces this to under a second on an Apple M2 CPU, enabling its use in retrieval pipelines. Generating the full dataset (30,000 samples) with the kinetic solver required approximately two weeks using 64 CPU cores, whereas generating the same number of samples with the trained network took around 26 minutes on a desktop computer.

Overall network performance is assessed using a combination of numerical metrics and visual evaluation. The GNN is evaluated against the U-Net model using the same dataset and test split. Table \ref{tab:metrics_summary} summarises a numerical comparison between the two architectures, demonstrating that the GNN model outperforms the U-Net model, with a mean MAE on the test set lower by approximately a factor of 3 than that of the U-Net model.

\begin{figure*}
\centering
\includegraphics[width=0.7\linewidth]{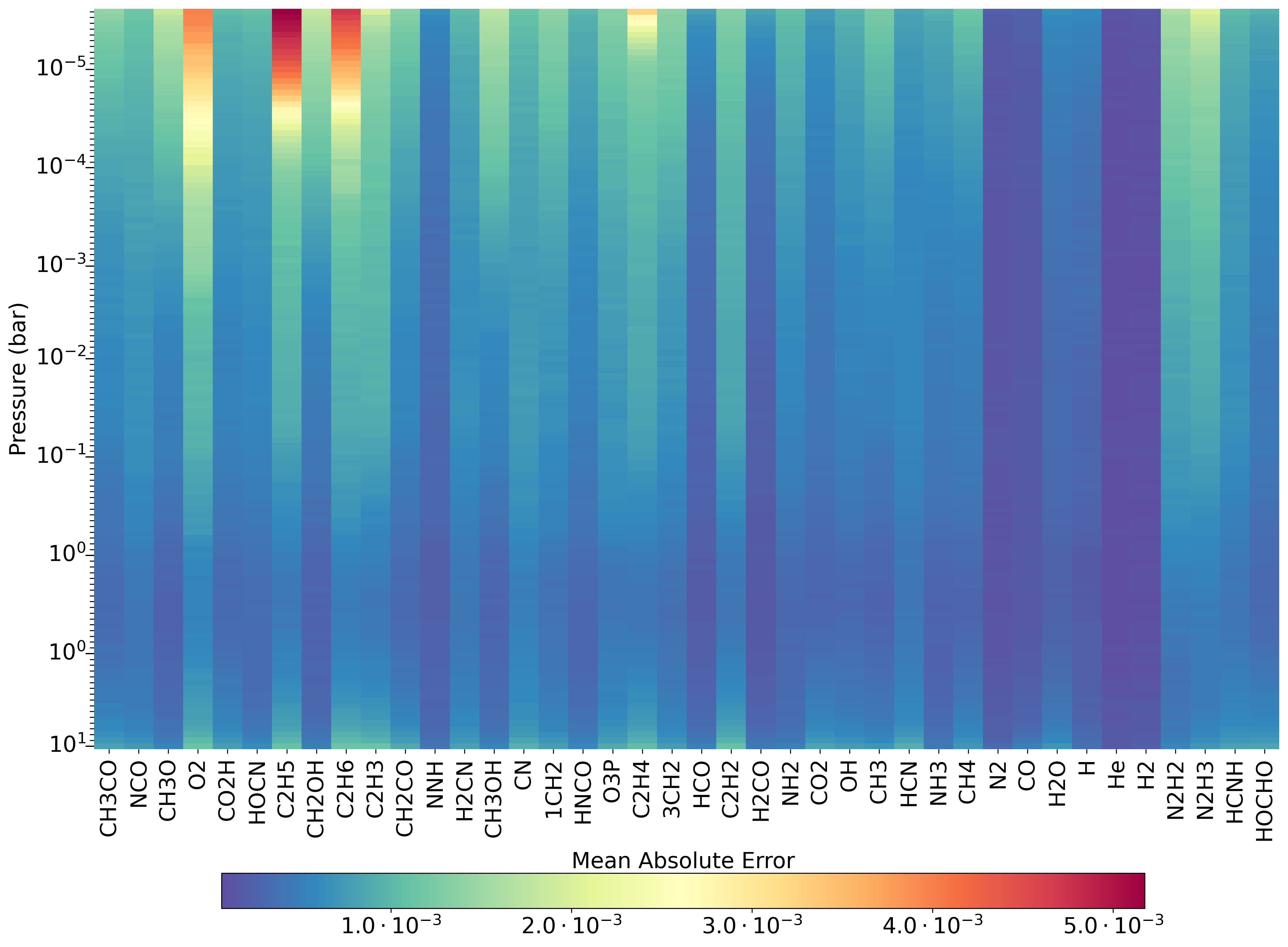} 
\caption{Mean absolute error of the test dataset per molecule and pressure layer of the trained GNN as a function of pressure for individual molecular species. The most significant errors are associated with three species (C$_2$H$_5$, C$_2$H$_6$, and O$_2$), which are affected by jitter removed in the pre-processing stage, where jittered profiles were interpolated. }
\label{fig:mean_heatmap}
\end{figure*}

Figure \ref{fig:mean_heatmap} presents the mean absolute error per chemical species for the GNN. The species with the highest errors (O$_2$, C$_2$H$_5$, C$_2$H$_6$) are those that require interpolation in low-abundance regimes, due to jitter caused by the FRECKLL code.  Aside from these cases, slightly elevated errors are observed at lower pressures. 

\begin{figure*}
\centering
\includegraphics[width=0.7\linewidth]{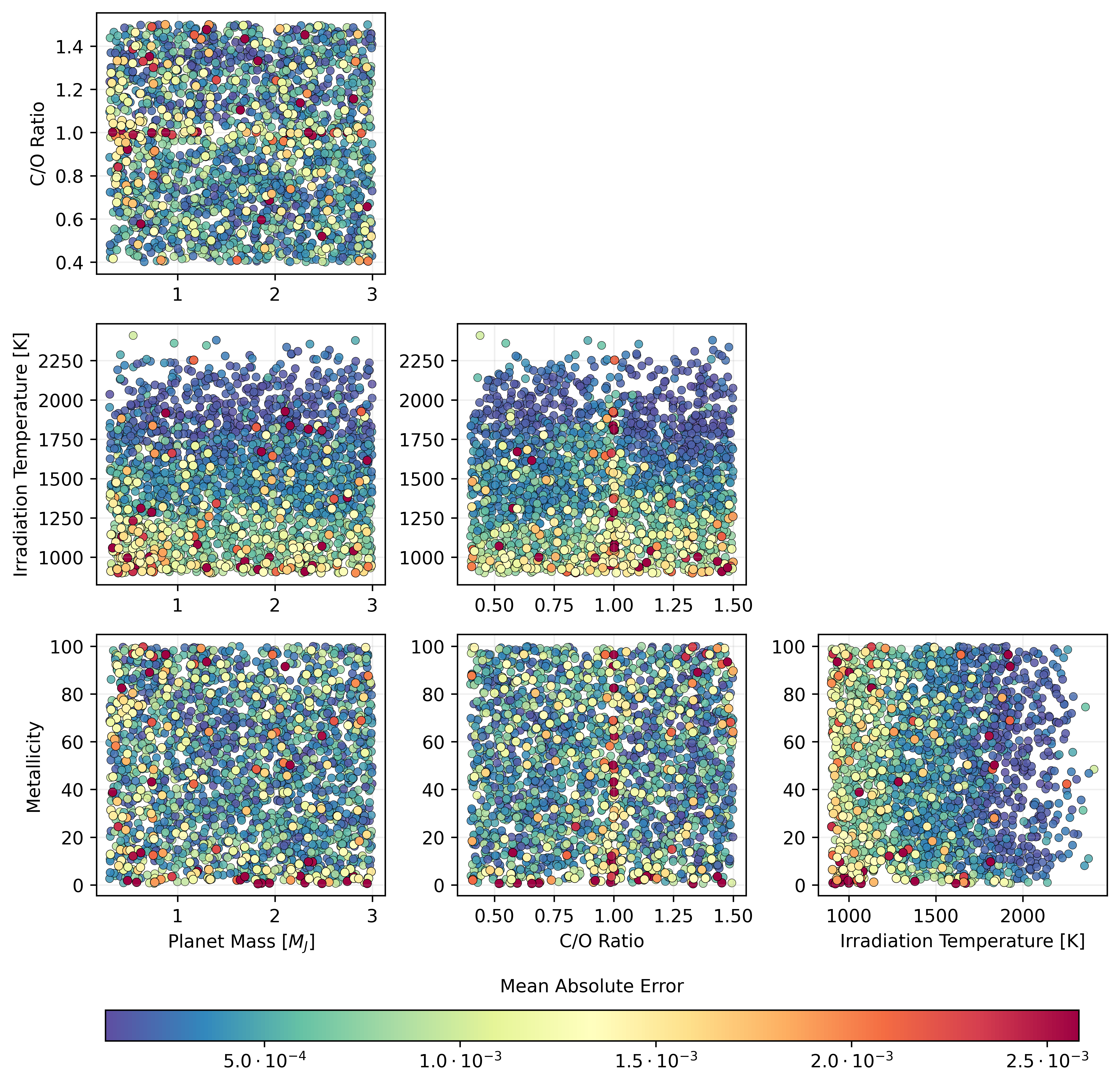} 
\caption{Mean absolute error per atmosphere as a function of the initial planetary parameters. 
Consistent with previous work \citep{vojtekova2025}, increased errors are observed for atmospheres with C/O ratios close to one and for lower irradiation temperatures. For visualisation purposes, the MAE values are clipped at the 99th percentile to suppress extreme outliers and enhance the visibility of underlying patterns.}
\label{fig:initial_parameters}
\end{figure*}

Figure \ref{fig:initial_parameters} provides a visualisation of the MAE per atmosphere given distributions of initial parameters. For visual clarity, MAE values were clipped at the 99th percentile.  Consistent with previous results reported in \citep{vojtekova2025}, we observe higher errors for samples with C/O ratios around one and at lower temperatures. 

The highest-error sample (referred to as atmosphere A1) evaluated in Table~\ref{tab:gnn_samples} and with initial parameters in Table~\ref{tab:samples_properties} had a C/O ratio of 1, which is a known challenging regime for atmospheric retrievals. In contrast, the sample at the 95th percentile of error performs substantially better, indicating that the worst-case sample is a statistical outlier rather than representative of the model's general behaviour. For a sample with an abundance MAE near the median, the model achieves a reduction in error of more than an order of magnitude compared to the worst sample, and the resulting spectral MAE is effectively undetectable with current observational capabilities.

\begin{figure*}
\centering
\includegraphics[width=0.49\linewidth]{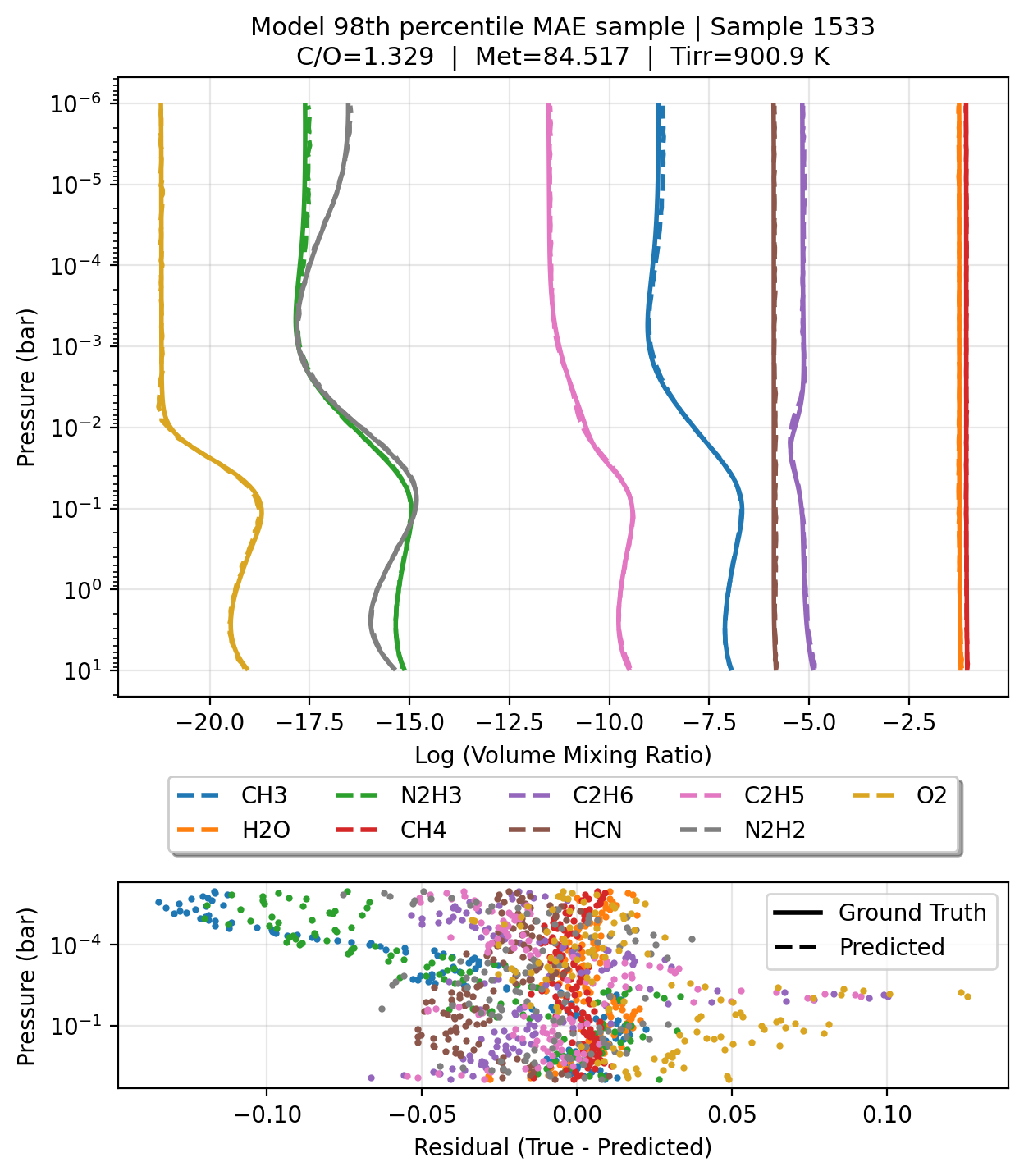} 
\includegraphics[width=0.49\linewidth]{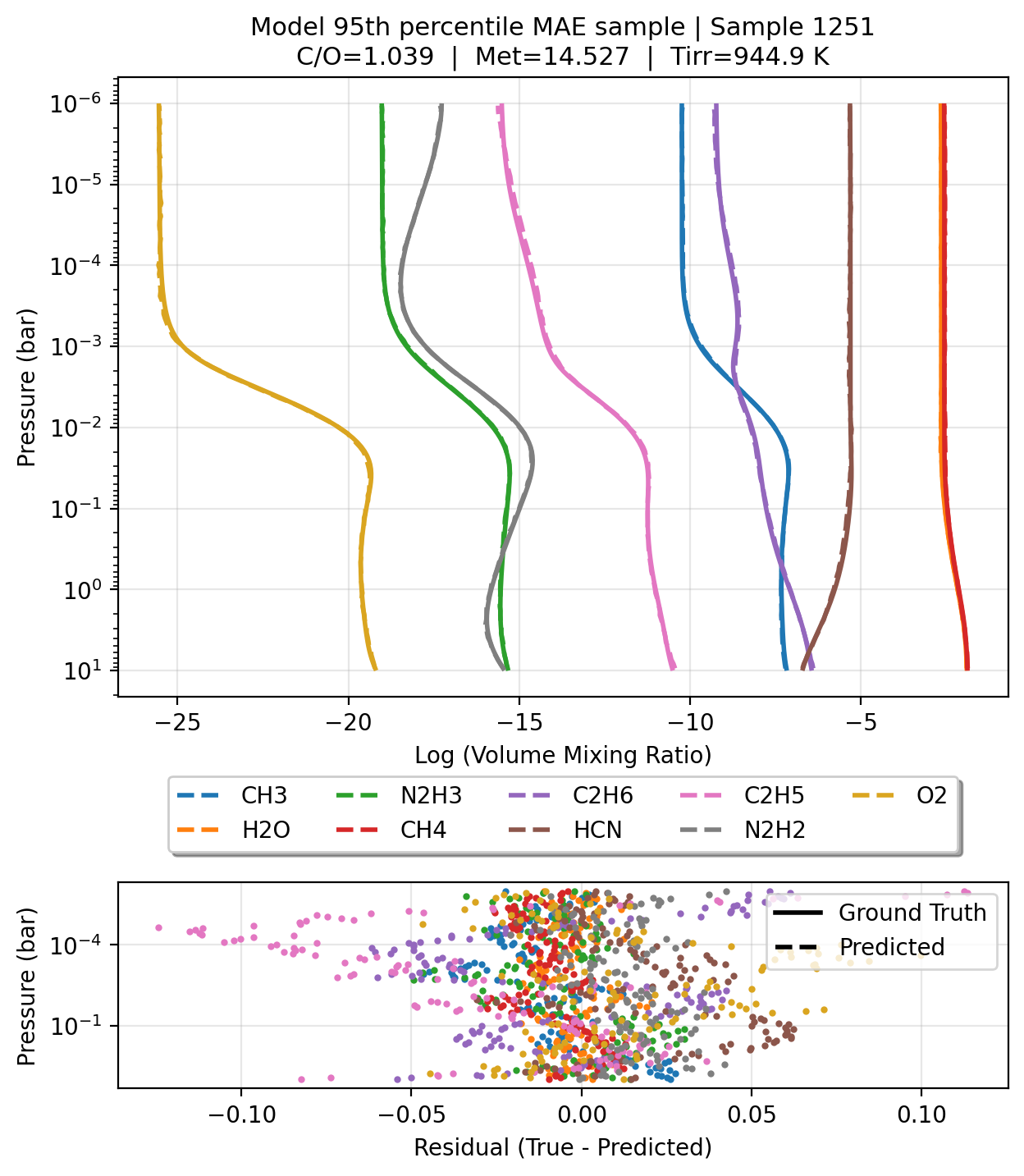} 
\caption{Reconstruction of vertical abundance profiles and corresponding residuals for two test atmospheres selected near the upper tail of the error distribution. The right plot shows a sample close to the 95th percentile of the mean absolute error (Atmosphere A2). The left plot shows a sample close to the 98th percentile. For each atmosphere, the upper panel shows the true (solid lines) and predicted (dashed lines) logarithmic volume mixing ratio profiles of selected molecular species, while the lower panel displays the corresponding residuals.}
\label{fig:sample_vmr_plots}
\end{figure*}

\begin{figure*}
\centering
\includegraphics[width=0.49\linewidth]{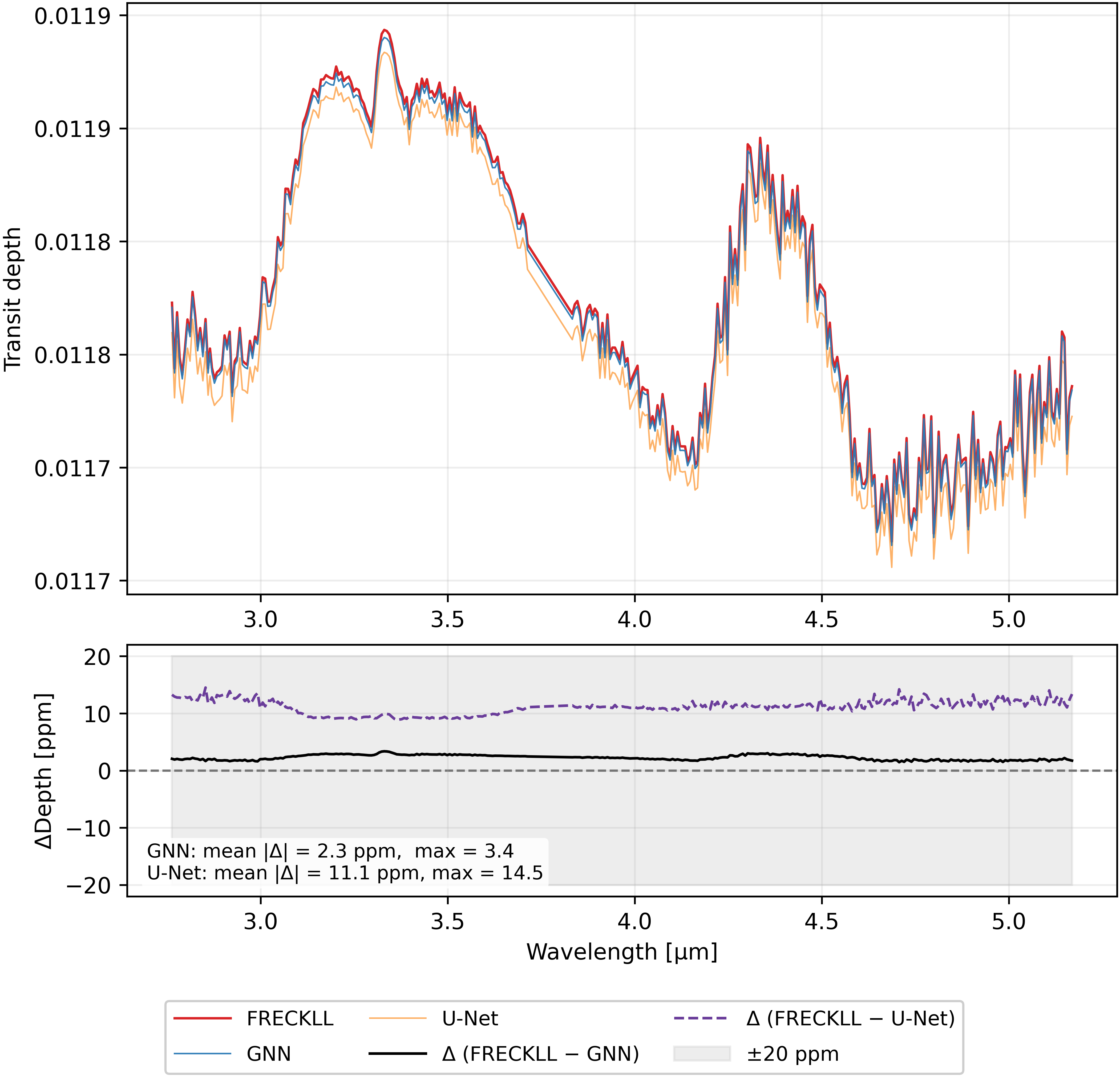} 
\includegraphics[width=0.49\linewidth]{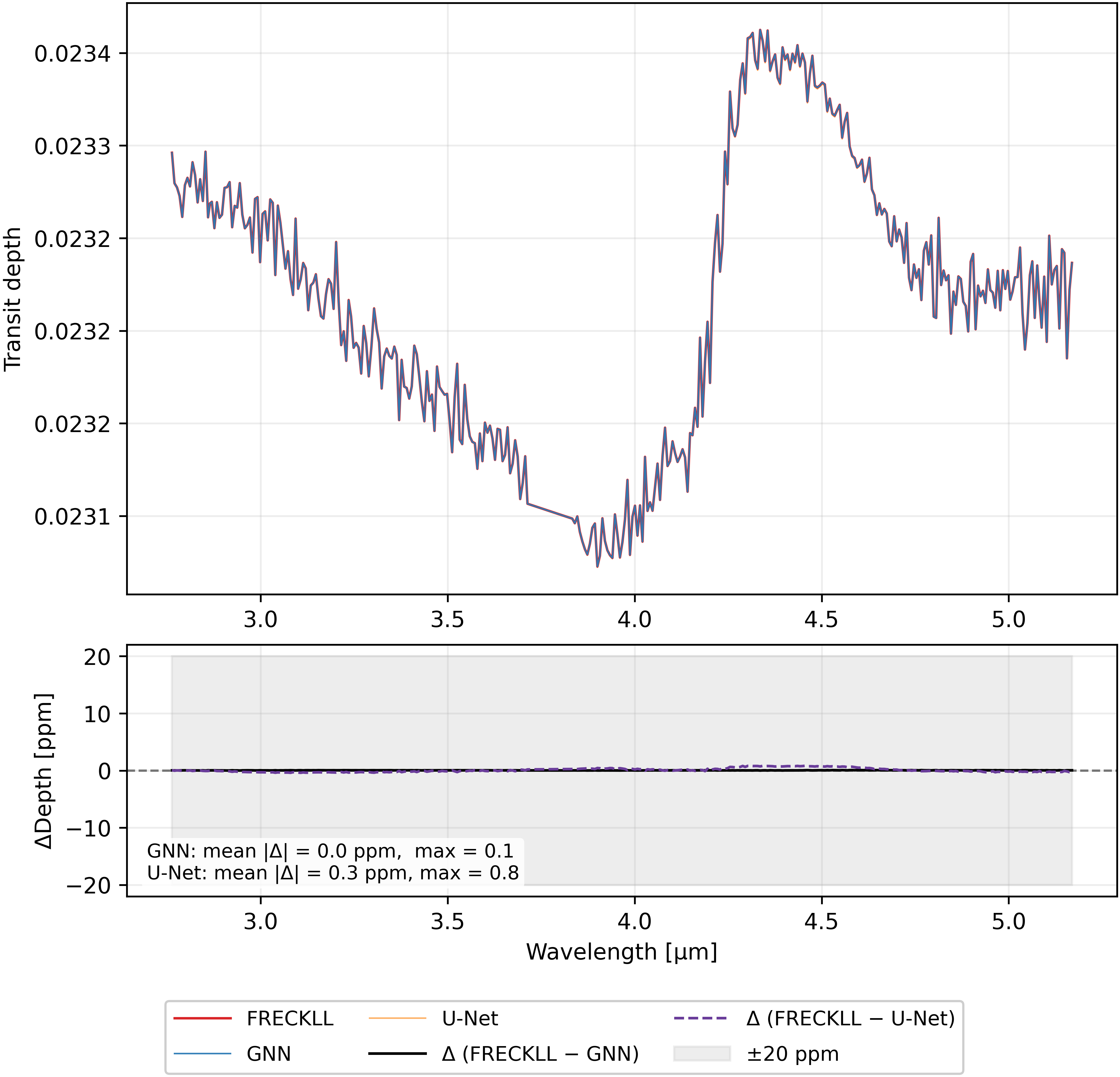}
\caption{Comparison between transmission spectra generated with FRECKLL (red), the GNN network (blue), and the U-Net network (orange), binned to the JWST/NIRSpec G395H resolution, for two representative atmospheres from the test dataset (Atmospheres A7 and A8). The upper panels show the binned transit depth as a function of wavelength, while the lower panels display residuals with respect to FRECKLL in parts per million (ppm). The shaded region indicates the ±20 ppm reference interval. On the left is atmosphere A7, selected as the atmosphere with the median error on the spectrum evaluation metric. The median-case example provides a representative case of typical model behaviour, avoiding emphasis on either especially favourable or especially challenging reconstructions. The right plot shows atmosphere A8, which corresponds to the spectrum with the lowest reconstruction error.}
\label{fig:spectrum_median_best}
\end{figure*}

\begin{figure*}
\centering
\includegraphics[width=0.49\linewidth]{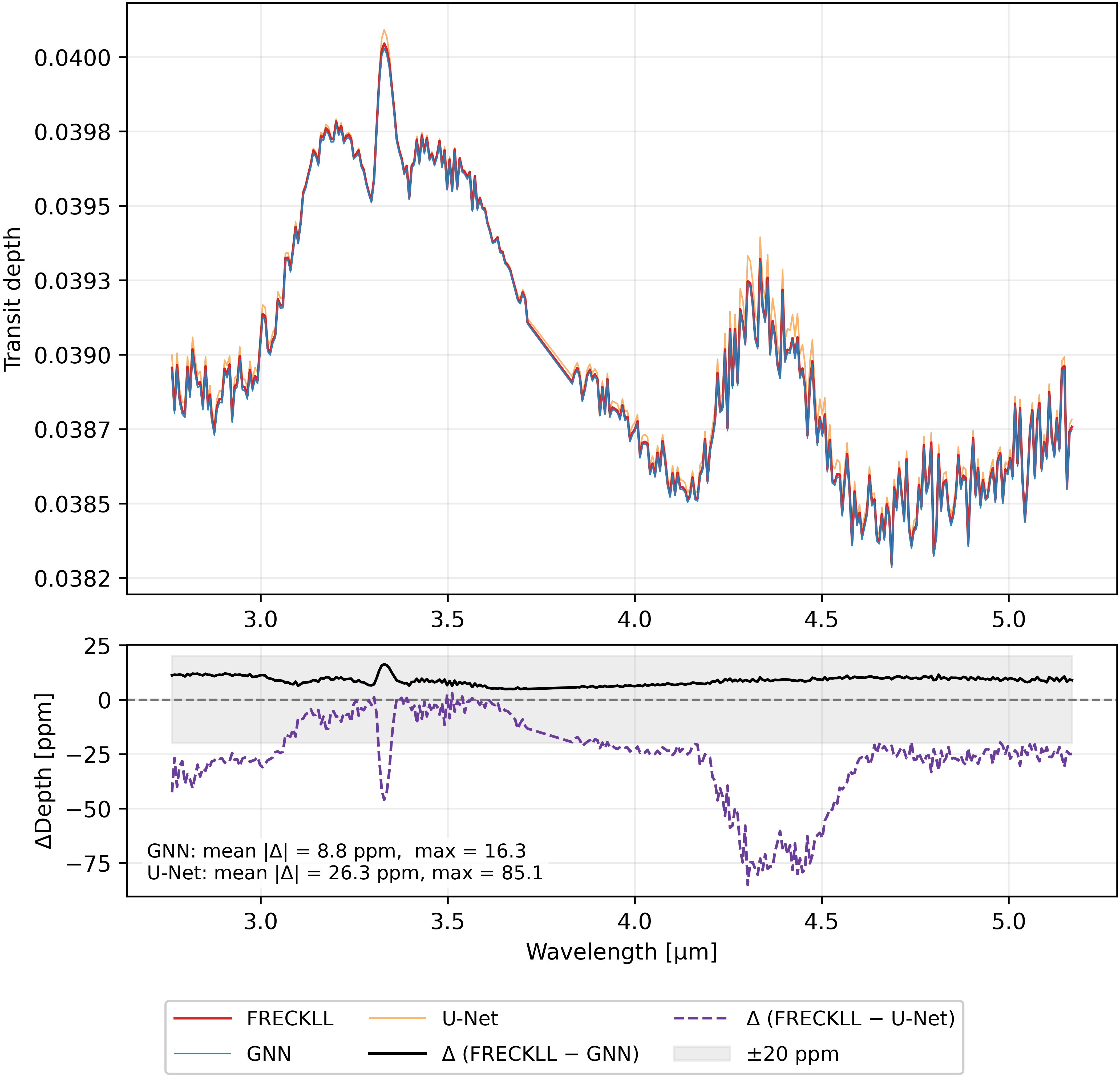} 
\includegraphics[width=0.49\linewidth]{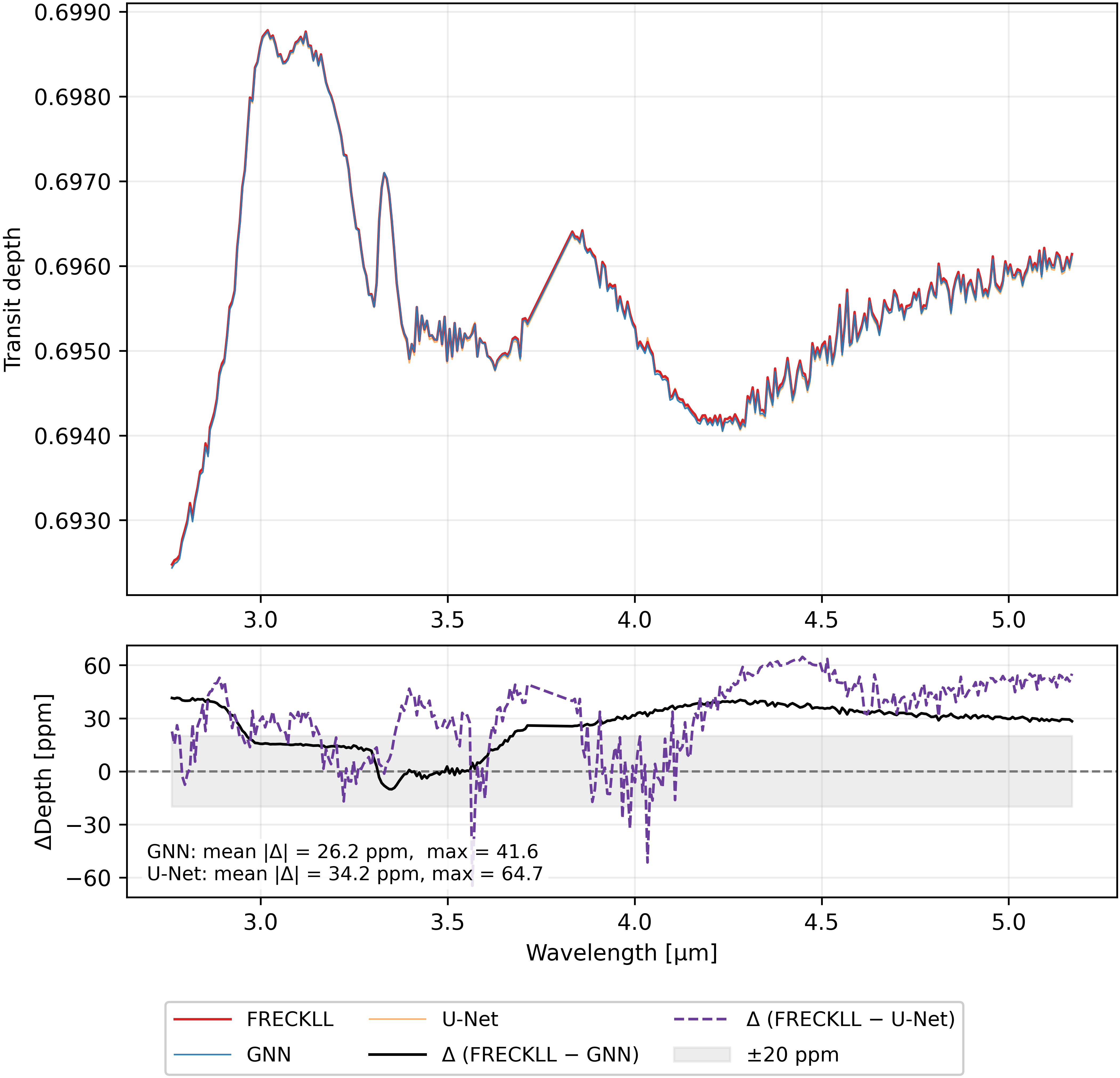}
\caption{Comparison between transmission spectra generated with FRECKLL (red), the GNN network (blue), and the U-Net network (orange), binned to the JWST/NIRSpec G395H resolution, for two representative atmospheres from the test dataset. The upper panels show the binned transit depth as a function of wavelength, while the lower panels display residuals with respect to FRECKLL in parts per million (ppm). The shaded region indicates the ±20 ppm reference interval. On the left is atmosphere A2, selected near the 95th percentile of abundance reconstruction error. The right plot shows the atmosphere A6 selected near the 95th percentile of spectral reconstruction error for GNN-generated samples. For the full unbinned spectra, the mean absolute error is $\approx$ 10 ppm (GNN) and $\approx$ 20 ppm (U-Net) in the left panel, and $\approx$ 28 ppm (GNN) and $\approx$ 48 ppm (U-Net) in the right panel.}
\label{fig:spectrum}
\end{figure*}

Figure~\ref{fig:sample_vmr_plots} illustrates representative atmospheric reconstructions corresponding to the 98th and 95th percentile error cases. While high-error atmospheres exhibit noticeable deviations in molecular abundances, converting these abundance errors into transmission spectra reveals that these discrepancies become minor when accounting for the observational uncertainties expected from JWST. This demonstrates that even significant MAE values in abundance predictions may have minimal impact on the spectra. This is because the spectral information primarily originates from mid-pressure atmospheric layers, whereas the largest abundance errors typically occur in low-pressure regions. Additionally, these low-pressure errors are influenced by the jitter described in the data section. Figure~\ref{fig:spectrum_median_best} shows the samples with the median and lowest reconstruction errors evaluated on spectra.
Figure~\ref{fig:spectrum} compares two representative samples: one with a 95th percentile error based on abundance data (left figure) and another with a 95th percentile spectral error (right figure). These plots also compare the GNN and U-Net results, showing that although the U-Net generally yields a higher mean error in the spectra, there is considerable sample-to-sample variability, and in some cases, the U-Net surpasses the GNN. Analysis of the spectra generated by the GNN reveals that about 7\% of test samples (168 out of the test set) have a mean absolute error (MAE) greater than 20 ppm, while roughly 2\% exceed 50 ppm.

\subsection{WASP-39b}
Evaluation on WASP-39b, with parameters near the boundary of the training distribution, demonstrates moderately reduced performance compared to in-distribution test atmospheres. The abundance reconstruction mean absolute error for this sample is $7.42 \cdot 10^{-3}$, which is above the 95th-percentile error (normalised-space MAE) of the synthetic test set (Atmosphere A2 - $1.43 \cdot 10^{-3} $).

\begin{figure*}
\centering
\includegraphics[width=0.6\linewidth]{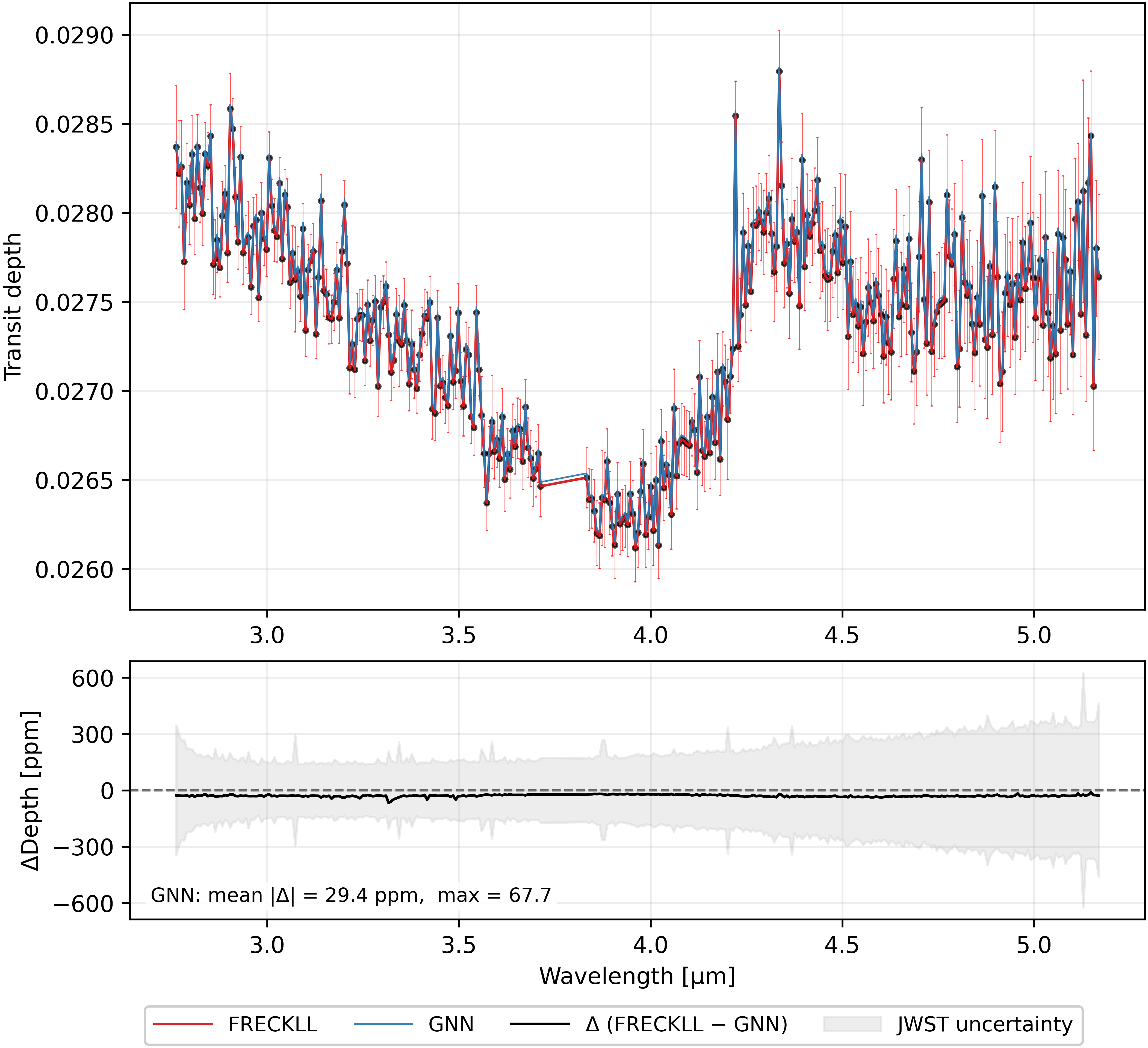} 
\caption{Transmission spectrum of WASP-39b predicted by the GNN (blue) compared with the FRECKLL reference (red). Bottom: residuals $\Delta$(FRECKLL - GNN) in ppm; grey band marks error bounds.}
\label{fig:spectrum_wasp39b}
\end{figure*}

The corresponding transmission spectrum (Figure \ref{fig:spectrum_wasp39b}) exhibits a systematic offset toward lower transit depths in the GNN prediction. The mean absolute error of the unbinned spectrum is 26 ppm.

Because WASP-39b represents an edge-of-distribution atmosphere in both temperature–pressure structure and planetary parameters, this degree of performance degradation aligns with the expected extrapolative behaviour of the emulator and remains within the upper-tail performance regime of the test distribution.

\subsection{Perturbation}
\begin{figure*}
\centering
\includegraphics[width=0.6\linewidth]{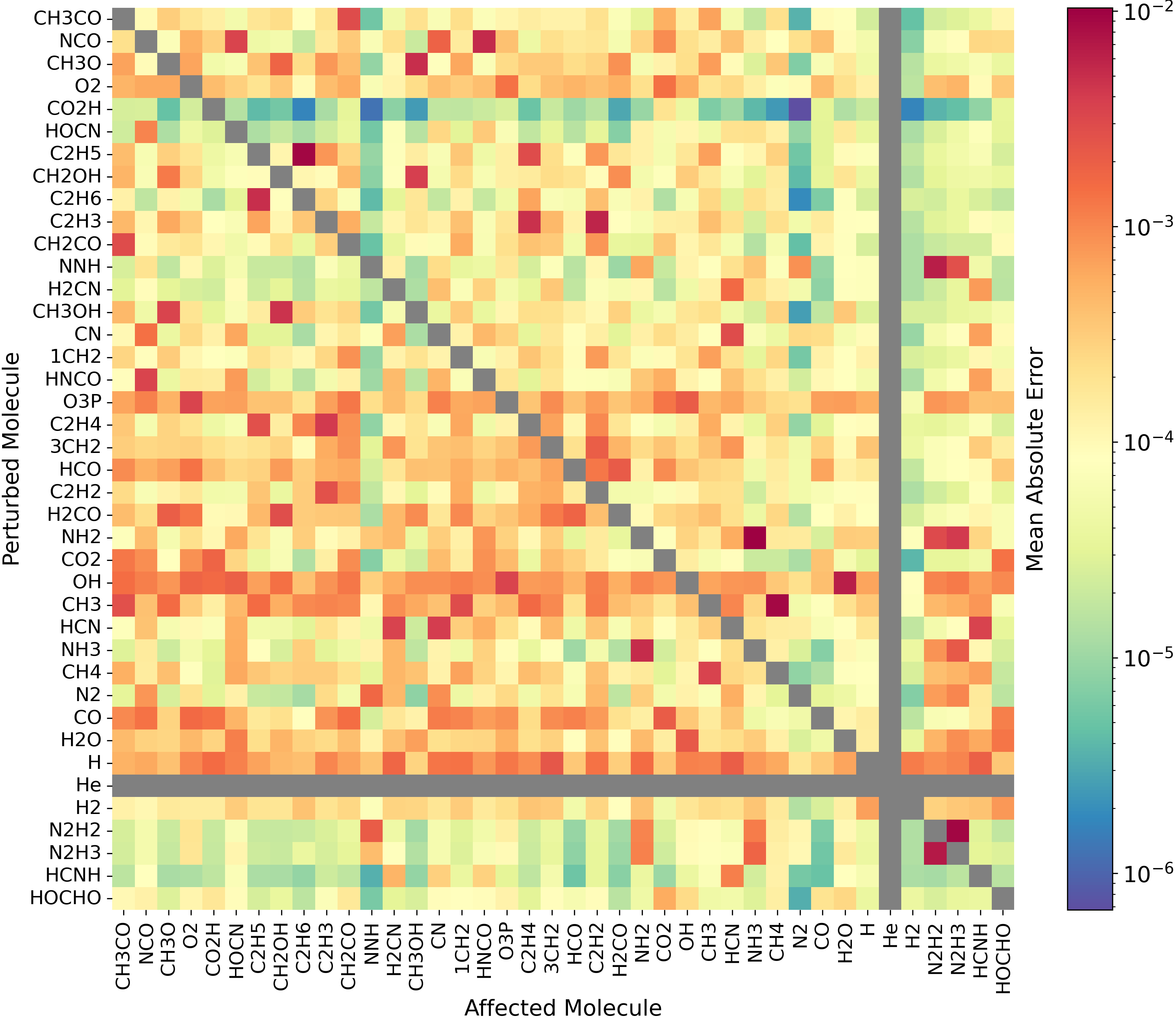} 
\caption{ 
Perturbation sensitivity matrix showing the mean absolute reconstruction error induced in each molecule (columns) when perturbing a given input molecule (rows). Colours indicate the magnitude of the resulting error, with red corresponding to the strongest disturbance, and the colour map is shown on a logarithmic scale to highlight structure across the full dynamic range of responses. The diagonal elements have been rescaled for visualisation purposes, as self-perturbations systematically produce the largest errors and would otherwise dominate the colour normalisation. He is connected only through a self-loop in the graph, so perturbing He affects only its own reconstruction; consequently, the corresponding row and column are shown in grey.}
\label{fig:perturbation}
\end{figure*}

Figure \ref{fig:perturbation} demonstrates that the propagated error does not exhibit diagonal clustering patterns, as observed with a convolutional neural network (CNN) in \cite{10.1093/rasti/rzag002}, Figure 11. This result suggests that the error propagates through molecular connections (edges), aligning with the intended effect of altering the network architecture.

This raises the question of whether perturbing a molecule primarily causes the disturbance to propagate along its strongest outgoing edge.

It was initially hypothesised that the most significant disturbances would propagate along the most abundant or strongest edges between species. However, subsequent analysis revealed that additional factors, such as the number of reactions in which a molecule participates, also influence the propagation of perturbations. The spread of disturbance depends not only on the strength of chemical edges but also on the number of other reactants with which the affected molecule interacts.

Based on this behaviour, fragile species are defined as those that are easily disturbed by perturbations from their connected neighbours, whereas steady species remain stable even when coupled to a perturbed molecule. This classification does not indicate molecular importance but rather characterises the manner in which information propagates within the network.

Fragile species, such as C$_2$H$_5$ and C$_2$H$_6$, consistently exhibit strong error responses when their most strongly connected chemical neighbour is perturbed. In contrast, steady species, including OH and H, are highly robust; they rarely appear among the highest-error targets, even when strongly coupled, due to their extensive connectivity.

For example, the species C$_2$H$_5$ connects to C$_2$H$_6$ through 28 edges, resulting in a strong error response. Its second strongest product, C$_2$H$_4$, shares 14 edges with C$_2$H$_5$ and produces the second strongest error response, as illustrated in Figure \ref{fig:perturbation}.

In contrast, the species NH$_2$ shares 30 connections with NH$_3$ and 17 with H; however, perturbing NH$_2$ does not create a strong error response in H. Another example is the species $^3$CH$_2$, which has its strongest and second-strongest connections to H (20 edges) and CO (9 edges), yet its perturbation results in the largest error in C$_2$H$_2$, with which it shares only five edges. This outcome is likely because H and CO receive information from a substantially larger number of species in the graph than C$_2$H$_2$.

Visualisation of Figure \ref{fig:perturbation} with linear normalisation highlights several off-diagonal responses that are more noticeable and clearly non-symmetrical. The most prominent pairs are:
\begin{itemize}
    \item C$_2$H$_5 \rightarrow$ C$_2$H$_6$,
    \item NH$_2 \rightarrow$ NH$_3$,
    \item CH$_3 \rightarrow$ CH$_4$,
    \item N$_2$H$_2 \rightarrow$ N$_2$H$_3$,
    \item OH $\rightarrow$ H$_2$O.
\end{itemize}
Two key questions emerge: why do these pairs produce stronger perturbation responses than others, and why are the responses non-symmetrical? 

The non-symmetrical behaviour appears to be related, at least in part, to differences in the connectivity of the affected species within the reaction network. In these prominent pairs, the more strongly disturbed species generally participates in fewer reactions than its counterpart, making it less buffered against perturbations and therefore more fragile in the learned representation.

Explaining why these particular pairs produce stronger responses than the rest is less straightforward and likely involves several factors acting together. 

First, all of these pairs are chemically closely related, most differing by one hydrogen atom, and therefore share closely linked pathways. This makes it plausible that the network learns a particularly strong local dependence between them. 

Second, the direct connections between these pairs are also relatively strong. Specifically, if we look at each reactant and retain only the product with which it shares the largest number of connections, and then sum the strengths of those connections, we find that four out of the five main perturbation hotspots are among the ten strongest reactant--product pairs in the network.

Finally, the response also depends on whether the affected molecule is itself relatively robust or fragile, since even a strong connection does not necessarily lead to a dominant perturbation if the target species is well buffered by the rest of the species in the graph. 

This interpretation remains incomplete, since the present analysis does not include the abundances themselves, their variation from sample to sample, the equilibrium-to-disequilibrium changes, or the baseline reconstruction error of individual species, all of which are also likely to influence the observed perturbation magnitudes. 

Additionally, this analysis does not account for temperature, which affects connection strength, or the specific initial parameters of samples. These factors are likely to influence the results, but a more detailed investigation is beyond the scope of this study.

\begin{table*}
\centering
\begin{tabular}{lccc|ccc}
\toprule
Metric &  \multicolumn{3}{c|}{GNN Model} &  \multicolumn{3}{c}{U-Net model} \\
  & Normalised data & log(VMR) & Spectrum [ppm] & Normalised data & log(VMR)  & Spectrum [ppm]\\
\midrule
MAE & \( 6.22 \cdot 10^{-4} \) & \( 9.33 \cdot 10^{-3} \)  & \(7.24\) & \( 1.60 \cdot 10^{-3} \) & \( 2.40 \cdot 10^{-2} \) & \( 16.22 \) \\
STD & \( 2.20 \cdot 10^{-3} \) & \( 3.30 \cdot 10^{-2} \)  & \(24.87\) & \( 4.64 \cdot 10^{-3} \) & \( 6.97 \cdot 10^{-2} \)  & \( 43.27 \)\\
95th percentile of MAE & \( 1.43 \cdot 10^{-3} \) &  \( 2.15 \cdot 10^{-2}\)  & \( 27.90\) & \( 4.43 \cdot 10^{-3} \) & \( 6.65 \cdot 10^{-2} \)  & \( 64.52 \)\\
Percentage > 50\% MAPE & 0.39\,\%  & - & -& 1.04\,\% & -  & -\\
Samples > 50\% MAPE & 9 & - & -& 24  & -  &-\\
\bottomrule
\end{tabular}
\caption{Comparison of predictive performance between the graph neural network and the U-Net architecture on the test set. Metrics are reported in the normalised space used for network training, in log(VMR), a standard metric in exoplanet atmospheric chemistry and on the generated spectrum in parts per million. }
\label{tab:metrics_summary}
\end{table*}

\begin{table*}
    \centering
\begin{tabular}{llccccccl}
\toprule
Atmosphere & Index & \multicolumn{3}{c}{MAE} & \multicolumn{3}{c}{STD} & Note \\
 &  & Norm & log(VMR) & Spectrum [ppm] &  Norm & log(VMR) & Spectrum [ppm]& \\
\midrule
A1 & 1276 & \( 1.78 \cdot 10^{-2} \) & \( 2.67 \cdot 10^{-1} \) & \(16.33 \) & \( 3.60 \cdot 10^{-2} \) & \( 5.40 \cdot 10^{-1} \) & \( 26.11 \) & Worst MAE \\
A2 & 1251 & \( 1.43 \cdot 10^{-3} \) & \( 2.15 \cdot 10^{-2} \) & \(9.96 \) & \( 2.05 \cdot 10^{-3} \) & \( 3.07 \cdot 10^{-2} \) & \( 3.10 \) & 95$^{th}$ percentile sample \\
A3 & 1674 & \( 4.67 \cdot 10^{-4} \) & \( 7.00 \cdot 10^{-3} \) & \( 2.59 \) & \( 3.46 \cdot 10^{-3} \) & \( 5.20 \cdot 10^{-2} \) & \( 6.00 \cdot 10^{-1} \) & Median sample \\
A4 & 2046 & \( 1.54 \cdot 10^{-3} \) & \( 2.31 \cdot 10^{-2} \) & \( 10.04 \) & \( 2.07 \cdot 10^{-3} \) & \( 3.10 \cdot 10^{-2} \) & \( 6.33 \) & Random sample \\
A5 & 1937 & \( 9.36 \cdot 10^{-5} \) & \( 1.40 \cdot 10^{-3} \) & \( 1.94 \) & \( 1.24 \cdot 10^{-4} \) & \( 1.87 \cdot 10^{-3} \) & \(3.24 \cdot 10^{-1} \) & Lowest MAE \\
A6 & 1686 & \( 4.42 \cdot 10^{-4} \) & \( 6.62 \cdot 10^{-3} \) & \(28.07 \)& \( 6.26 \cdot 10^{-4} \) & \( 9.39 \cdot 10^{-3} \) & \( 10.83\) & 95$^{th}$ spectrum percentile sample \\
A7 & 2171 & \( 8.79 \cdot 10^{-4} \) & \( 1.32 \cdot 10^{-2} \) & \(1.86 \)& \( 1.17 \cdot 10^{-3} \) & \( 1.75 \cdot 10^{-2} \) & \( 1.09\) & Median spectrum sample \\
A8 & 868 & \( 4.13 \cdot 10^{-4} \) & \( 6.19 \cdot 10^{-3} \) & \( 6.29 \cdot 10^{-2} \)& \( 6.48 \cdot 10^{-4} \) & \( 9.72 \cdot 10^{-3} \) & \( 6.99 \cdot 10^{-2}\)  & Best spectrum sample \\
\bottomrule
\end{tabular}
\caption{Detailed evaluation of selected atmospheric samples predicted by the GNN model. 
The table reports representative cases, including the worst-performing, 95th-percentile, median, lowest-error, and a randomly selected atmosphere. For each sample, error metrics are provided in the normalised space, in log(VMR) units and in generated spectra. The corresponding atmospheric parameters are listed in Table~\ref{tab:samples_properties}, and the associated T–P profiles are shown in Figure ~\ref{fig:TP_profiles}.}
\label{tab:gnn_samples}
\end{table*}

\begin{table*}
    \centering
\begin{tabular}{lcccccccccccc}
\toprule
Atm. & $R_{\mathrm{p}}\,$ & $M_{\mathrm{p}}\,$ & Met. & $\mathrm{C/O}$ & $T_{\mathrm{irr}}\,$ & $T_{\mathrm{int}}\,$ & $\alpha$ & $\kappa_{\mathrm{IR}}$ & $\kappa_{\mathrm{v1}}$ & $\kappa_{\mathrm{v2}}$ & $\gamma_1$ & $\gamma_2$ \\

 &$[R_{\mathrm{J}}]$  &  $[M_{\mathrm{J}}]$&  & $\mathrm{Ratio}$ & $[\mathrm{K}]$ & $[\mathrm{K}]$ &  & &  & & & \\
\midrule
A1 & $ 1.52 $ & $ 2.34 $ & $ 9.7 $ & $ 1.00 $ & $ 1814 $ & $ 93 $ & $ 3.8 \cdot 10^{-1} $ & $ 4.8 \cdot 10^{-2} $ & $ 3.6 \cdot 10^{-2} $ & $ 3.9 \cdot 10^{-2} $ & $ 7.5 \cdot 10^{-1} $ & $ 8.2 \cdot 10^{-1} $ \\
A2 & $ 1.25 $ & $ 0.51 $ & $ 14.5 $ & $ 1.04 $ & $ 945 $ & $ 111 $ & $ 5.0 \cdot 10^{-1} $ & $ 4.2 \cdot 10^{-2} $ & $ 1.1 \cdot 10^{-2} $ & $ 5.7 \cdot 10^{-3} $ & $ 2.7 \cdot 10^{-1} $ & $ 1.4 \cdot 10^{-1} $ \\
A3 & $ 1.00 $ & $ 0.62 $ & $ 83.2 $ & $ 1.40 $ & $ 1582 $ & $ 64 $ & $ 4.2 \cdot 10^{-1} $ & $ 1.9 \cdot 10^{-2} $ & $ 9.3 \cdot 10^{-3} $ & $ 8.7 \cdot 10^{-3} $ & $ 5.0 \cdot 10^{-1} $ & $ 4.7 \cdot 10^{-1} $ \\
A4 & $ 1.82 $ & $ 0.72 $ & $ 96.5 $ & $ 0.62 $ & $ 980 $ & $ 124 $ & $ 6.9 \cdot 10^{-1} $ & $ 1.1 \cdot 10^{-2} $ & $ 5.4 \cdot 10^{-3} $ & $ 8.5 \cdot 10^{-3} $ & $ 4.9 \cdot 10^{-1} $ & $ 7.7 \cdot 10^{-1} $ \\
A5 & $ 1.35 $ & $ 2.30 $ & $ 53.7 $ & $ 1.20 $ & $ 2015 $ & $ 78 $ & $ 6.3 \cdot 10^{-1} $ & $ 4.8 \cdot 10^{-2} $ & $ 2.4 \cdot 10^{-2} $ & $ 1.3 \cdot 10^{-2} $ & $ 5.1 \cdot 10^{-1} $ & $ 2.8 \cdot 10^{-1} $ \\
A6 & $ 1.30 $ & $ 1.68 $ & $ 96.6 $ & $ 1.19 $ & $ 1387 $ & $ 101 $ & $ 3.7 \cdot 10^{-1} $ & $ 5.2 \cdot 10^{-2} $ & $ 4.6 \cdot 10^{-2} $ & $ 1.3 \cdot 10^{-2} $ & $ 8.9 \cdot 10^{-1} $ & $ 2.4 \cdot 10^{-1} $ \\
A7 & $ 1.01 $ & $ 0.46 $ & $ 89.5 $ & $ 1.11 $ & $ 960 $ & $ 105 $ & $ 4.0 \cdot 10^{-1} $ & $ 6.0 \cdot 10^{-2} $ & $ 1.1 \cdot 10^{-2} $ & $ 2.0 \cdot 10^{-2} $ & $ 1.8 \cdot 10^{-1} $ & $ 3.4 \cdot 10^{-1} $ \\
A8 & $ 1.32 $ & $ 2.34 $ & $ 76.9 $ & $ 0.90 $ & $ 1512 $ & $ 57 $ & $ 5.4 \cdot 10^{-1} $ & $ 3.8 \cdot 10^{-2} $ & $ 1.6 \cdot 10^{-2} $ & $ 2.7 \cdot 10^{-2} $ & $ 4.1 \cdot 10^{-1} $ & $ 7.1 \cdot 10^{-1} $ \\
\bottomrule
\end{tabular}
\caption{Atmospheric and planetary parameters of the selected samples analysed in Table~\ref{tab:gnn_samples}. Listed parameters include the planetary radius and mass, metallicity, C/O ratio, irradiation and internal temperatures, and the temperature--pressure profile parameters used to construct the atmospheric models.}
\label{tab:samples_properties}
\end{table*}
\section{Discussion}
The primary motivation for modifying the architecture was to address the inductive spatial bias introduced by the two-dimensional convolutional kernel in a U-Net-like architecture. The graph neural network architecture enables the encoding of actual chemical connections between species, thereby eliminating inductive bias and facilitating information propagation consistent with physical and chemical principles. Unlike convolutional architectures that assume Euclidean neighbourhood structure, the GNN operates on the non-Euclidean topology of chemical reaction networks, allowing information exchange along physically meaningful reaction pathways rather than grid-adjacent spatial locations. Additionally, the GNN offers greater flexibility for capturing complex Guillot temperature–pressure profiles. This study demonstrates a progression from generic machine learning (U-Net) to physically informed machine learning (GNN). While both approaches yield satisfactory accuracy, the GNN reduces structural bias and enhances interpretability for scientific applications. 

Consistent with \cite{vojtekova2025}, persistent error patterns emerge, especially near a carbon-to-oxygen (C/O) ratio of 1 and at low temperatures. These results support the interpretation that such errors stem from physically meaningful challenges, rather than random failures or architectural limitations.

Elevated errors observed in species such as C$_2$H$_5$, C$_2$H$_6$, and O$_2$ are primarily due to numerical artefacts in the FRECKLL solver and the interpolation methods used during preprocessing, not inherent limitations of the machine-learning architecture.

Beyond the initial network evaluation, spectral assessment was performed by converting the network outputs into transmission spectra. This method is effective because transmission spectra are most sensitive to mid-pressure layers, making deviations in upper- and lower-atmosphere abundances less apparent. The assessment shows that spectral errors remain moderate for JWST and anticipated Ariel data, although occasional outliers are observed. To further mitigate observational errors, a practical strategy involves using surrogate models to explore the broad parameter space during retrievals, then transitioning to a kinetic code for final results as convergence is approached.

To evaluate the network under edge-case conditions, spectra were predicted for WASP-39b, which possesses a temperature–pressure profile near the boundary of the training distribution. The model's performance aligns with evaluation errors, falling near the 95th percentile. Although the mean absolute error (MAE) in chemical predictions is in the higher tail of the error distribution, the spectrum reconstruction error remains below 30 ppm, a value that is scarcely detectable in actual observations, if we account for JWST standard errors for such observations, which are at the level of 50 -- 100 ppm. This test demonstrates the surrogate model's effectiveness under conservative and realistic conditions. The observed degradation near this boundary case is consistent with expected surrogate behaviour outside densely sampled regions of parameter space and does not indicate instability of the learned chemical mapping.

Beyond standard evaluation, the propagation of perturbations through the network was analysed and compared with that of the previous U-Net model. Perturbations predominantly propagate along chemically meaningful connections, though connectivity alone does not wholly determine the response; the degree of a species’ involvement in chemical reactions also affects its robustness or susceptibility. While the relationship is complex, it aligns with the network’s architectural intent. These tests affirm that the architecture effectively encodes chemical topology, a behaviour distinct from that of convolutional architectures, where perturbations propagate mainly through spatial kernel overlap rather than along chemically connected species.

The forward modelling test demonstrates that the GNN accurately reconstructs spectra, with errors consistent with expectations for JWST and Ariel, and achieves a speed-up of more than an order of magnitude relative to kinetic codes. This improvement enables the inclusion of disequilibrium chemistry in retrieval analyses.
Limitations of this study include the exclusive use of Guillot temperature–pressure profiles, a fixed K$_{zz}$, and the restricted boundaries of the training set. In this work, the model was trained on the Venot+2020 C/H/O/N chemical scheme. As chemical schemes continue to evolve, with updated or extended schemes such as \cite{2024A&A...682A..52V, 2026A&A...706A.260V}, the surrogate would need to be retrained and revalidated to remain physically consistent. Future work should focus on integrating the network directly into the retrieval pipeline. Furthermore, photochemistry or complex cloud-condensation models were not included in the present model but can be incorporated in future iterations.

\section{Conclusions}

The calculation of disequilibrium steady-states abundances remains a major computational challenge in atmospheric retrieval pipelines, and the growing precision of exoplanet observations increasingly invalidates the assumption of chemical equilibrium. Although the previously introduced U-Net surrogate achieved satisfactory accuracy, it introduced an artificial spatial bias inconsistent with the chemical reaction network topology. This study describes a graph neural network surrogate that encodes real chemical connections between species and reactions, trained on atmospheres with Guillot temperature–pressure profiles and Venot+2020 chemical scheme. The results show that the GNN accurately reconstructs disequilibrium abundances across the sampled parameter space. When mapped to transmission spectra, the average errors remain within the observational precision expected for JWST and Ariel. Performance variations are observed in chemically transitional regimes, notably near a C/O ratio of one and at low temperatures, consistent with the behaviour observed in the U-Net model. Compared with the convolutional architecture, perturbation analysis shows improved structural fidelity, with disturbances propagating along reaction connectivity rather than spatial adjacency. Evaluation of the boundary case WASP-39b further indicates robustness under moderate domain shift.

The accurate results make it feasible to include a disequilibrium chemistry surrogate model in the retrieval pipeline for JWST and Ariel, with the advantage of providing orders-of-magnitude computational acceleration relative to kinetic codes.

\section*{Acknowledgements}
We thank Jack Davey for helpful additional information regarding the binning of JWST-like spectra.
A. Vojtekova acknowledges funding from ESA via the OSIP platform with contract number 4000139240.
O.V. acknowledges funding from the ANR project `EXACT' (ANR-21-CE49-0008-01) and from the Centre National d'Etudes Spatiales (CNES).

\section*{Data Availability}
The official GitHub repository for this study is currently private and will be released publicly upon publication of the paper. Details of the network architecture are included in the supplementary material.




\appendix
\section{Hyperparameter Search}
Table \ref{tab:hyperparameter_search} provides an overview of the hyperparameter space explored for the main architectural components. Model selection was performed iteratively. Configurations that demonstrated unstable training or exhibited high training and validation losses were excluded. A subset of stable, well-performing architectures was retrained using multiple fixed random seeds to evaluate robustness. The final model was selected from these stable candidates based on performance on the test set.

\begin{table*}
\centering
\caption{Hyperparameter search space explored during development of the graph neural network. The table summarises the principal architectural and training hyperparameters investigated, together with the final configuration adopted. The search was performed iteratively rather than as a fully parameter space sweep.}
\label{tab:hyperparameter_search}
\begin{tabular}{llll}
\hline
Hyperparameter & Tested options & Final choice & Note \\
\hline
Global encoder hidden dimension & 128, 256 & 256 & Hidden width of the global encoder MLP \\
Edge encoder hidden dimension & 128, 256 & 256 & Hidden width of the edge encoder MLP \\
Node encoder hidden dimension & 128, 256, 512 & 512 & Hidden width of the node encoder MLP \\
Decoder hidden dimension & 128, 256, 512, 1024 & 1024 & Hidden width of the decoder MLP \\
Global encoding dimension & 64, 128 & 128 & Size of output encoded global feature vector \\
Number of graph layers & 2, 3, 4, 5 & 4 & Message-passing depth \\
Aggregation function & sum, mean & sum & Aggregation of neighbouring messages \\
Batch size & 4, 8, 16 & 8 & Training batch size \\
Learning rate & $10^{-3},\,10^{-4}$ & $10^{-4}$ & Initial learning rate \\
\hline
\end{tabular}
\end{table*}





\bsp	
\label{lastpage}
\end{document}